\documentclass[aps,pre,twocolumn,groupedaddress]{revtex4-1}

\usepackage[pdftex]{graphicx}
\usepackage{amssymb,amsfonts,amsmath}

\newcommand{\BibTeX}{\textsc{Bib}\TeX}       

\begin{document}


\title{Efficient hierarchical analysis of the stability of a network through dimensional reduction of its influence topology}


\author{Ali Kinkhabwala}
\affiliation{Department of Systemic Cell Biology, Max Planck Institute of Molecular Physiology, Otto-Hahn-Str.~11, 44227 Dortmund, Germany; tel.: +49-231-1332222; email: ali.kinkhabwala@mpi-dortmund.mpg.de}


\date{\today}

\begin{abstract}
The connection between network topology and stability remains unclear. General approaches that clarify this relationship and allow for more efficient stability analysis would be desirable. Inspired by chemical reaction networks, I demonstrate the
utility of expressing the governing equations of arbitrary first-order dynamical systems (interaction networks) in terms of sums of real functions (generalized \textit{reactions}) multiplied by real scalars (generalized \textit{stoichiometries}). Specifically, I examine the mathematical notion of influence topology, which is fundamentally based on the network's so-defined reaction stoichiometries and the first derivatives of the reactions with respect to each species at the steady state solution(s). The influence topology is naturally represented as a signed directed bipartite graph with arrows or blunt arrows connecting a species node to a reaction node (positive/negative derivative) or a reaction node to a species node (positive/negative stoichiometry). The set of all such graphs is denumerable. A significant reduction in dimensionality is possible through stoichiometric scaling, cycle compaction, and temporal scaling. All cycles in a network can be read directly from the graph of its influence topology, enabling efficient and intuitive computation of the principal minors (sums of products of non-overlapping bipartite cycles) and the Hurwitz determinants (sums of products of either the principal minors or the bipartite cycles) for testing steady state stability. The stability of a given network is shown to have a hierarchical dependence first on its influence topology and then, more specifically, on algebraic conditions (exact functional form of the reactions). The utility of this hierarchical approach to bifurcation analysis is demonstrated on classical networks from control theory, biology, chemistry, physics, and electronics. Due to its fundamental nature and denumerability, the influence topology provides a useful tool for systematic characterization of the connection between network topology and stability.
\end{abstract}


\maketitle

\section{Introduction}

The formal study of dynamical systems was initiated by Poincar\'e \cite{strogatz_nonlinear_1994}, who, among his many contributions, importantly introduced in 1885 the notion of a \textit{bifurcation} \cite{poincare_equil_1885}, which denotes a dramatic alteration in the dynamical properties of a system upon a change in one or more of the parameters that define it. Bifurcation theory is largely concerned with algebraic classification of the critical values of appropriately defined bifurcation parameters \cite{strogatz_nonlinear_1994,kuznetsov_elements_2004}. In recent years, this classical approach has been augmented by the introduction of powerful group theoretic tools for examination of previously unappreciated symmetries of the governing equations and phase space solutions of many important classes of dynamical systems \cite{golubitsky_nonlinear_2006, gilmore_symmetry_2007}. The presence of these symmetries leads to an often significant simplification of the description of the system's dynamics (akin to dimensional reduction). In this manuscript, I introduce a novel and completely general form of local bifurcation analysis that is based on recognition of the fundamental nature of the topological cycles (and their symmetries) that arise from consideration of dynamical systems as interaction networks. This unique viewpoint stems from the synthesis of several disparate notions associated with the diverse fields of chemistry (generalized reactions/stoichiometries, cycles), control theory (Routh-Hurwitz conditions, Routh array root counting), mathematics (combinatorics, topology, graph theory), and physics (dimensional reduction of the parameters that define the network). Further elaboration of the connection of the properties of a network's \textit{bipartite} cycles as presented below (including their symmetries and their implications for dimensional reduction of the problem of local stability) with classical bifurcation theory \cite{strogatz_nonlinear_1994,kuznetsov_elements_2004} and with the seemingly orthogonal symmetries corresponding to a network's governing equations and their phase space solutions (including their group properties and implications for dimensional reduction) \cite{golubitsky_nonlinear_2006, gilmore_symmetry_2007} represents an interesting challenge for the future.

Topological approaches for understanding dynamical systems (interaction networks) come in two flavors: the study of the topology of the network's possible phase space trajectories (\textit{phase space topology}, for which Poincar\'e also famously laid the foundations \cite{stillwell_mathematics_2010}) and the distinct study of the topological structure of a network's interactions (\textit{interaction topology}). In this manuscript, I will use the word ``topology'' only in the latter sense of referring to the structure of a network's interactions. Such topological approaches have had a long yet relatively sparse history of successful application to real-world networks, beginning with Kirchhoff's current and voltage laws from 1847 \cite{kirchhoff_ueber_1847}, which were significantly generalized by Weyl in 1923 \cite{weyl_reparticion_1923} and, independently, Tellegen in 1952 \cite{tellegen_general_1952,tellegen_general_1953,penfield_tellegens_1970}. Another important example is Stueckelberg's 1952 proof that quantum mechanical unitarity already entails bilateral normalization, cyclic recurrence, and Boltzmann's $H$ theorem \cite{stueckelberg_theoreme_1952,watanabe_symmetry_1953,gorban_michaelis-menten-stueckelberg_2011}.

Over the past half century, topological methods have come to play an important role in particular for the study of chemical reaction networks, with the diverse pioneering studies of King \& Altmann \cite{king_schematic_1956}, Bak \cite{bak_contributions_1963}, Morowitz et al.~\cite{morowitz_passive_1964}, Higgins \cite{higgins_theory_1967}, Hill \cite{hill_thermodynamics_1968}, Gardner \& Ashby \cite{gardner_connectance_1970}, and others from the 50's and 60's anticipating the more systematic approaches of the next decade, for example, the ``network thermodynamics'' of Oster, Perelson, \& Katchalsky \cite{oster_network_1971,perelson_network_1975}. In 1972, Horn \& Jackson \cite{horn_general_1972} and Feinberg \cite{feinberg_complex_1972} importantly proved that the unipartite graphs that convey a chemical network's \textit{complex topology}, in which reactant complex nodes are connected to product complex nodes through transition edges, can be used to address the number and stability of the steady states of certain classes of mass action networks \cite{feinberg_existence_1995,feinberg_multiple_1995,craciun_understanding_2006}. This ultimately master equation-based approach was already implicit in Stueckelberg's proof mentioned above; for example, Feinberg's ``zero deficiency theorem'' \cite{feinberg_complex_1972} is a direct consequence of Stueckelberg's more general proof based on the network's \textit{configuration topology} \cite{gorban_michaelis-menten-stueckelberg_2011}. In a contemporaneous yet orthogonal direction, Vol'pert in 1972 \cite{volpert_differential_1972}, and independently Clarke in 1974 \cite{clarke_graph_1974}, introduced the formal study of the bipartite graphs that convey a chemical reaction network's \textit{stoichiometric topology}, with separate nodes for the species and reactions and with edges displaying reactant or product stoichiometries. 
This early work, along with further work in the 70's and 80's by Vol'pert and his collaborators \cite{volpert_analysis_1985,volpert_mathematical_1987,mincheva_graph-theoretic_2006} and by Clarke \cite{clarke_stability_1980}, collectively demonstrated the usefulness of the stoichiometric topology for examination of steady state multiplicity and stability in particular for mass action networks. A contemporaneous and highly reductionist approach based on a network's unipartite \textit{logical topology} (or ``kinetic logic''), which displays the positive/negative effect of one species on another according to the original governing equations, was first explored by Glass \& Kauffman in 1973 \cite{glass_logical_1973} and later by Thomas \cite{thomas_ed._kinetic_1979} and King \cite{king_kinetic_1987}. Another minimal and completely general perspective originally inspired by economic models was proposed by Quirk, Ruppert, \& Maybee in the late 60's \cite{quirk_qualitative_1965,maybee_qualitative_1969,maybee_combinatorially_1974} and later comprehensively addressed by Jeffries \cite{jeffries_qualitative_1974,jeffries_when_1977}. This approach involves examination of a network's sign stability, requiring knowledge only of the signs of the terms in the network's Jacobian matrix obtained upon first-order perturbation of the steady state (\textit{sign topology}). Clarke examined some of the implications of sign stability for chemical reaction networks in 1980 \cite{clarke_stability_1980}; see Do et al. \cite{do_graphical_2012} for more recent results on the general problem of sign stability. The overly simplified perspectives utilized by both the sign and logical topologies, however, prohibit more detailed examinations of network stability. It is important to note that the strongest and most useful results from all of the above studies on chemical networks were obtained only under the assumption of mass action kinetics, which restricts the reaction functions to positive products of reactant concentrations raised to stoichiometric exponents.

\begin{figure}[]
\vspace{-4.5cm}
\centerline{\hspace{0cm}     \includegraphics*[width=0.5\textwidth]{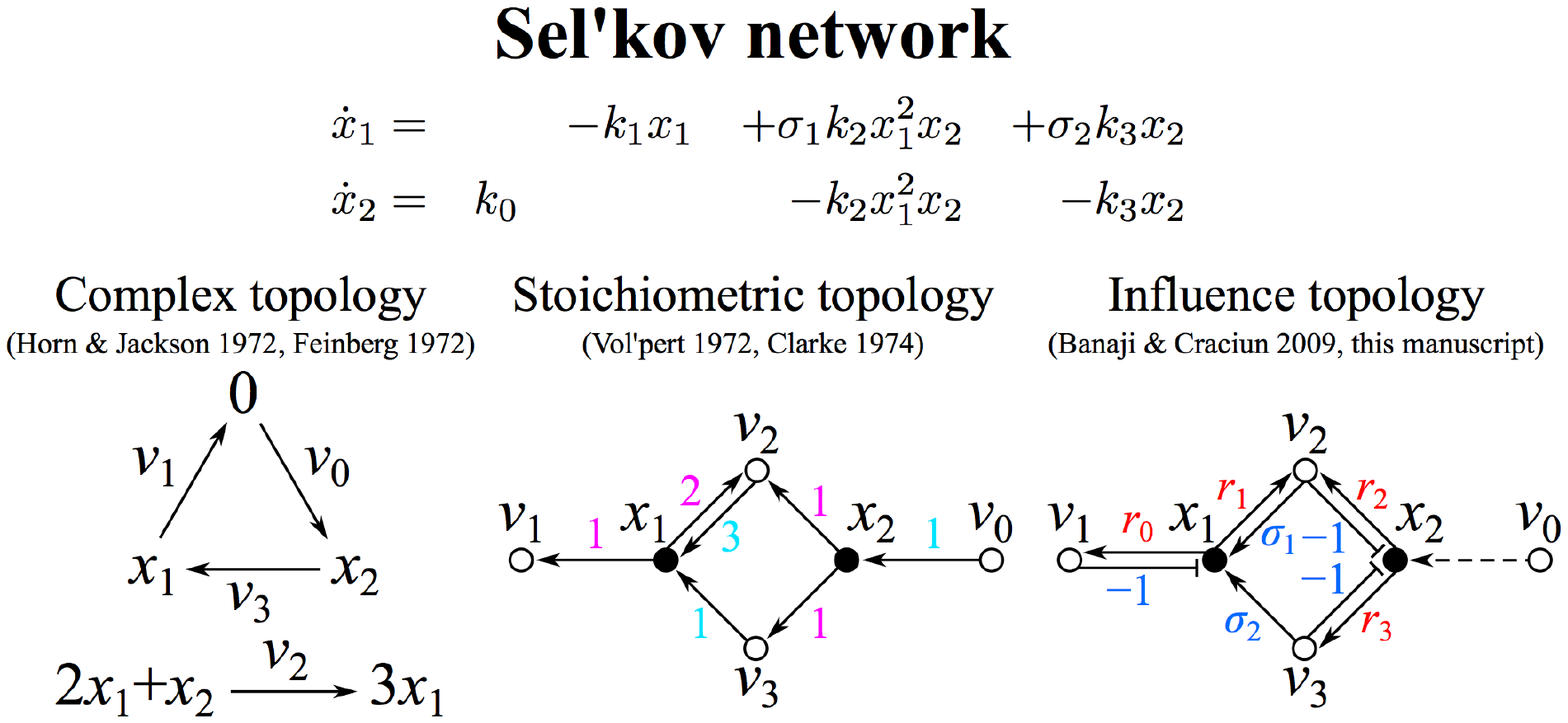}}
\vspace{-4.5cm}
      \caption{Different topological representations of the Sel'kov network. For the traditional network, $\sigma_1=\sigma_2=1$, however these coefficients are needed to define the influence topology, which represents a generalized form of the original network. The $i$ subscript in the $k_i$ parameters of the governing equations labels the different reactions $v_i$ that appear in the topological graphs. The edges in the stoichiometric topology are labeled with the reactant (magenta) or product (cyan) stoichiometries. The edges in the dimensionally reduced influence topology are labeled with Jacobian (red) or stoichiometric (blue) parameters.}
      \label{fig:topologies}
\end{figure}

In the above summary, five different topological interpretations of chemical networks (and, in some cases, more general networks) have been mentioned based on a network's sign, logical, configuration, complex, and stoichiometric topologies.
In this manuscript, a distinct and completely general interpretation is explored based on a network's perturbative \textit{influence topology}, which naturally emerges from examination of the first order perturbation of a network steady state. The influence topology corresponds to a signed directed bipartite graph with a \textit{Jacobian} edge from a species node to a reaction node encoding whether the reaction function increases (arrow) or decreases (blunt arrow) upon an increase in the species in the vicinity of the steady state (positive/negative reaction derivative with respect to the species), and a \textit{stoichiometric} edge from a reaction node to a species node encoding the positive (arrow) or negative (blunt arrow) sign of the species-specific stoichiometric factor for the reaction in the original governing equations (positive/negative stoichiometric coefficient). The complex, stoichiometric, and influence topologies are displayed in Fig.~\ref{fig:topologies} for the Sel'kov network \cite{selkov_self_1968}, which is used to model glycolytic oscillations (and will be further analyzed below). Chemists will immediately recognize the complex topology and stoichiometric topology, as these are the two principal ways in which chemical reaction networks are traditionally presented. The influence topology represents a completely general yet still intuitive mathematical structure. Here, a directed path from species $i$ to $j$ through reaction $k$ consists of a Jacobian edge followed by a stoichiometric edge, with their product yielding the net positive/negative contribution of reaction $k$ to the growth rate of species $j$ upon perturbation of species $i$ in the vicinity of the steady state. Unlike the more familiar complex and stoichiometric topologies, the usefulness of the influence topology is not restricted to mass action kinetics or even to the non-negativity of the species or reactions. As shown in detail below, the influence topology provides a simple and powerful calculational tool with which the local stability of \textit{arbitrary} first-order networks can be determined. It importantly reveals the hierarchical influence of first topology and then algebra on a network's local stability, with significant dynamical restrictions already imposed at the topological level. In the following sections, a general derivation of the influence topology is given followed by its detailed application to the study of local steady state stability for several classical networks.

Upon finalizing this manuscript, I discovered that Banaji and Craciun in 2009 \cite{banaji_graph-theoretic_2009,banaji_graph-theoretic_2010} had already introduced the influence topology --- which they refer to as Directed Species Reaction (DSR) graphs --- for the orthogonal purpose of studying the number of steady states of reaction networks. In very recent work, Angeli et al. examine various stability criteria that lead to algebraic and/or topological rules for exclusion of the appearence of instabilities within certain classes of networks \cite{angeli_combinatorial_2013}; however, they do not analyze the completely general and fundamental stability criteria provided by the Routh-Hurwitz conditions \cite{routh_treatise_1877,hurwitz_ueber_1895,gantmacher_applications_1959,meinsma_elementary_1995,jury_j.j._1996}. The implications of the Routh-Hurwitz conditions for chemical networks were already recognized by Bak \cite{bak_contributions_1963} and later examined in depth, both algebraically and topologically, for mass action networks by Clarke \cite{clarke_graph_1974,clarke_stability_1980}, with Wilhelm \cite{wilhelm_analysis_2007} more recently emphasizing their  general nature. In this manuscript, I demonstrate that the dimensionally reduced influence topology provides an efficient calculational tool for direct study of these fundamental criteria for completely arbitrary networks. Due to this novel perspective, most of the results presented below appear here for the first time, including  (1) several intuitive and highly useful notational conventions (both algebraic and graphical), (2) proof that only non-overlapping \textit{bipartite} cycles contribute to the principal minors, (3) topological expression of the Routh-Hurwitz criteria in terms of products of non-overlapping and overlapping cycles to help identify critical structures, (4) demonstration of dimensional reduction through cycle compactions and topological symmetries, (5) introduction of the stability phase space over the remaining dimensions of the problem to visualize the zones over which individual Hurwitz determinants are negative, (6) utilization of the Routh array to determine the exact number of unstable eigenvalues in each zone (and zone overlap), (7) identification of the set of \textit{fundamental} influence topologies with the set of all signed directed bipartite graphs that are constructed solely from interconnected cycles and that are not sign degenerate, and (8) illustrative examination of several classical networks from diverse fields.

Unlike most of the prior work mentioned above, the results obtained below require only very basic knowledge of linear algebra and ordinary differential equations, making this important field of study accessible to a wider audience. Biologists, chemists, physicists, and engineers should be able to immediately apply the simple and intuitive topological rules derived below for computation and visualization of the fundamental Routh-Hurwitz stability conditions for their specific networks of interest. The rich topological and algebraic structures revealed below --- complementing previous results on DSR graphs \cite{banaji_graph-theoretic_2009,banaji_graph-theoretic_2010,angeli_combinatorial_2013} --- should also be of interest to mathematicians.

\section{Dynamic interaction network}
Consider the following arbitrary system of autonomous first-order ordinary differential equations: 
\begin{equation}
\frac{dx_j}{dt}   =    f_j(x_1,\ldots,x_n),
\label{eq:ODE}
\end{equation}
with $j$ ranging from 1 to $n$. The $f_j(x_1,\ldots,x_n)$ denote real-valued functions of the real variables $x_j$.  This very general definition encompasses many important interaction networks studied in control theory, biology, chemistry, physics, and electronics \cite{strogatz_nonlinear_1994}.

\section{Steady state perturbation} 
Setting all $\dot{x}_j$ to zero allows determination of the one or more steady state solutions of the system through solution of the system of equations $f_j(x_1,\ldots,x_n)=0$. Perturbation about a particular steady state solution $(x_1^s,\ldots,x_n^s)$ yields:
\begin{align}
\frac{d(x^s_j+\Delta x_j)}{dt}   &=    f_j(x_1^s+\Delta x_1,\ldots,x_n^s + \Delta x_n) \nonumber\\
\frac{dx^s_j}{dt}+\frac{d\Delta x_j}{dt}   &=   f_j(x_1^s,\ldots,x_n^s) \nonumber\\
& \hspace{1.1em} +  \sum_i \Delta x_i \left(\frac{\partial f_j}{\partial x_i}\right)_s +\mathcal{O}\left((\Delta x)^2\right).
\end{align}
By the definition of the steady state, the first terms on the left- and right-hand sides are zero, yielding to first order:
\begin{equation}
\frac{d\Delta x_j}{dt}  \simeq   \sum_i \Delta x_i H_{ij},
\label{eq:perturb}
\end{equation}
with $H_{ij}\equiv(\partial f_j/\partial x_i)_s$ the ``transition rate constants'' from $i$ to $j$ defined at the steady state $s$. 
Considering $\mathbf{\Delta x}$ as a row vector (the index order is more intuitive, as we will see below), this can be expressed in matrix form as:
\begin{equation}
\frac{d}{dt}\mathbf{\Delta x}  \simeq  \mathbf{\Delta x\cdot H}.
\label{eq:matrixform}
\end{equation}

\section{Signs of the eigenvalues}
The stability of the steady state is determined by the signs of the real parts of the eigenvalues $\lambda_i$ of the associated eigenvectors $ \mathbf{(\Delta x)}_i\equiv\mathbf{L}_i$ of $\mathbf{H}$, which are defined by the equation $\lambda_i\mathbf{L}_i  = \mathbf{L}_i\mathbf{\cdot H}$ or
\begin{equation}
\mathbf{L}_i\mathbf{\cdot} \mathbf{(}\lambda_i\mathbf{I}-\mathbf{H} \mathbf{)} = \mathbf{0}.
\label{eq:eigenvalue}
\end{equation}
For a steady state to be stable, the real parts of all eigenvalues should be negative. 
For non-zero eigenvectors ($\mathbf{L}_i\neq\mathbf{0}$), Eq.~\ref{eq:eigenvalue} will only be true for singular (non-invertible) $\mathbf{(}\lambda_i\mathbf{I}-\mathbf{H} \mathbf{)}$ having
\begin{equation}
|\lambda\mathbf{I}-\mathbf{H}|=\rho(\lambda)=0.
\label{eq:determinant}
\end{equation}
In geometrical terms, the non-zero perturbation eigenvector $\mathbf{L}_i$ is ``perpendicular'' to $\mathbf{(}\lambda_i\mathbf{I}-\mathbf{H} \mathbf{)}$, with the latter matrix (which is a collection of column vectors) spanning only a subspace of the full $n$ dimensional space. The roots of the characteristic polynomial $\rho(\lambda)$ determine the eigenvalues and therefore the stability of the steady state:
\begin{equation}
\rho(\lambda)=a_0 \lambda^n+a_1 \lambda^{n-1}+a_2 \lambda^{n-2}+\cdots+a_{n-1} \lambda+a_n=0.
\label{eq:characteristic}
\end{equation}
While the first coefficient is equal to 1 by definition (Eq.~\ref{eq:determinant}), we will retain the notation $a_0$ below for clarity.  The coefficients $a_k$ can be expressed as:
\begin{align}
a_k&=\frac{1}{(n-k)!}\left(\frac{\partial^{n-k} \rho}{\partial \lambda^{n-k}}\right)_{\lambda=0}\nonumber\\
&=\frac{1}{(n-k)!}\left(\frac{\partial^{n-k}    }{\partial \lambda^{n-k}}  |\lambda\mathbf{I}-\mathbf{H}| \right)_{\lambda=0}.
\label{eq:ak}
\end{align}

\section{Principal minors as products of non-overlapping unipartite cycles}
From Eq.~\ref{eq:ak}, it is clear that $a_n=|-\mathbf{H}|=(-1)^n|\mathbf{H}|$. With a bit more effort, Eq.~\ref{eq:ak} can be shown to entail:
\begin{equation}
a_{q}=(-1)^{q}b_{q},
\end{equation}
with $b_q$ representing the $q\times q$ principal minor of $\mathbf{H}$ (with $b_0\equiv a_0$). 
Using the Leibniz rule, the principal minors of $\mathbf{H}$ can be written as:
\begin{equation}
b_q    =  \sum_{i_1<\cdots<i_q} \sum_{\pi(j_1,\ldots,j_q)} \epsilon_{j_1\ldots j_q}H_{i_1j_1}\cdots H_{i_qj_q},
\label{eq:Leibniz}
\end{equation}
with $\pi(j_1,\ldots,j_q)$ denoting the permutations of the ordered set $\{i_1,\ldots,i_q\}$ and $\epsilon_{j_1\ldots j_q}$ the Levi-Civita permutation symbol, equal to +1 for $j_1=i_1,\ldots,j_q=i_q$ and otherwise equal to $-1$ for odd permutations or $+1$ for even permutations. 

The specific presentation of the network perturbation in Eq.~\ref{eq:matrixform} as a species row vector perturbation multiplied on the right by the first-order transition matrix $\mathbf{H}$, permits a convenient reading of the unipartite (species-only) \textit{cycles} of the network from products of the $H_{ij}$ (with $i$ as usual denoting the row and $j$ denoting the column of $\mathbf{H}$). For example, $H_{12}H_{23}H_{31}$ corresponds to a 3-cycle from $1\rightarrow2$ then $2\rightarrow3$ then $3\rightarrow1$.  In the more traditional formalism consisting of multiplication of a column vector of perturbations by the transition matrix from the left (see Clarke \cite{clarke_graph_1974} for chemical networks or any textbook on quantum mechanics), these terms are $H_{21}H_{32}H_{13}$. In this traditional form, when reading the ordered transition matrix elements from left to right, one must confusingly read each paired subscript from right to left.

The principal minors have a simple topological interpretation as sums of products of non-overlapping cyclic permutations \cite{carmichael_introduction_1937} or, simply, cycles $c_l$ (with $l$ the number of species in the cycle):
\begin{align}
b_1 &= \sum_i H_{ii}=c_1,\notag\\
b_2 &= \frac{1}{2!}\sum_{i\neq j} H_{ii}H_{jj}-H_{ij}H_{ji} = \overline{c_1c_1}  -  c_2,\notag\\
b_3 &= \frac{1}{3!}\sum_{i\neq j,i\neq k, j\neq k} H_{ii}H_{jj}H_{kk}\notag\\
&\hspace{1.1em}-H_{ii}H_{jk}H_{kj}-H_{ik}H_{jj}H_{ki}-H_{ij}H_{ji}H_{kk}\notag\\
&\hspace{1.1em}+H_{ij}H_{jk}H_{ki}+H_{ik}H_{ji}H_{ki}=\overline{c_1c_1c_1} - \overline{c_1c_2}+c_3,\notag\\
&\hspace{0.5em}\vdots\notag\\
b_q&=\sum_{\substack{0\leq p_1\leq\ldots\leq p_q\\ \sum_i p_i = q}} (-1)^{s(p_1,\ldots, p_q)} \overline{c_{p_1}c_{p_2}\ldots c_{p_q}}.
\label{eq:minor}
\end{align}
Each principal minor, $b_q$, corresponds to the sum of the product of all \textit{unique} non-overlapping combinations of cycles (with cycle lengths summing to $q$) in the network. The cycle term, $c_l$, when appearing alone in the above expressions (i.e. without an overline), simply corresponds to the sum of all $l$-cycles in the network. The bar on top of the cycles indicates the non-overlapping nature of the unipartite cycles in the product, i.e. each species can only appear at most once in a particular cycle product. For example, $\overline{c_1c_2}$ represents the sum of all unique,  non-overlapping combinations of a 1-cycle and a 2-cycle in the network. In the final line, the $c_0\equiv1$ are simply placeholders and $s(p_1,\ldots, p_q)$ is a function that returns the number of (non-zero) even length cycles present in the list $\{p_1,\ldots,p_q\}$. In even simpler terms, $b_q$ is the sum of all non-overlapping cycle-based partitions of $q$, with a negative sign accompanying each even cycle in a given partition product. This topological definition of the principal minors is far more elegant and intuitive than the increasingly cumbersome index-based notation also displayed in the above for $b_1$, $b_2$ and $b_3$ (which is, not surprisingly, often misstated in the literature, e.g. \cite{wilhelm_analysis_2007}).

\section{Routh-Hurwitz stability conditions}
Negativity of one of the $a_q=(-1)^q b_q$ is sufficient to generate instability as can be easily seen from expression of the characteristic polynomial in terms of a product of its $R$ real roots and $I$ pairs of imaginary roots:
\begin{align}
\rho(\lambda) &= \prod_{r=1}^R(\lambda-\lambda_r)\prod_{i=1}^I(\lambda-\lambda_i)(\lambda-\bar{\lambda}_i)\notag\\
\rho(\lambda) &= \prod_{r=1}^R\left(\lambda+\left(-\lambda_r\right)\right)\prod_{i=1}^I\left(\lambda^2+2\left(-Re(\lambda_i)\right)+|\lambda_i|^2\right).
\end{align}
If all $\lambda_r$ and $Re(\lambda_i)$ are negative, all coefficients $a_q$ in Eq.~\ref{eq:characteristic} will clearly be positive; conversely, if one or more of the $a_q$ are negative, then at least one of the eigenvalues must be positive, implying that the steady state is unstable. While $a_q>0$ for all $q$ is clearly necessary for stability, it is not \textit{sufficient}. The Routh-Hurwitz conditions \cite{routh_treatise_1877,hurwitz_ueber_1895,gantmacher_applications_1959}, which are mathematically equivalent to the original criteria formulated by Hermite and the related criteria embodied in Lyapunov's second method \cite{hermite_number_1856,lyapunov_general_1992,parks_new_1962,parks_comment_1966,parks_new_1977,jury_j.j._1996}, provide both necessary and sufficient conditions for steady-state stability. While these conditions have traditionally been obtained through the use of sophisticated mathematics, a remarkably simple proof has been found more recently that requires only basic algebra and continuity arguments \cite{meinsma_elementary_1995}. The Routh-Hurwitz conditions for a stable steady state can be defined as:
\begin{equation}
\Delta_q>0
\end{equation}
for $q=1,\ldots,n$, with $\Delta_q$ denoting the Hurwitz determinant of the following matrix of the coefficients of the characteristic polynomial:
\begin{align}
\Delta_q=&
\left\lvert
\begin{array}{cccccc}
a_1  & a_0  & 0       & 0              &  \cdots  &   0 \\
a_3  & a_2  & a_1  & a_0         &  \cdots  &   0 \\ 
a_5  & a_4  & a_3  & a_2    &  \cdots  &   0 \\
a_7  & a_6  & a_5  & a_4  &  \cdots &    0 \\
\vdots & \vdots & \vdots & \vdots  &  \ddots & \vdots \\
a_{2q-1} & a_{2q-2} & a_{2q-3} & a_{2q-4} &  \cdots   & a_q
\end{array}
\right\rvert \notag\\
=&\left\lvert
\begin{array}{cccccccc}
-b_1  & b_0  & 0           & 0 &    \cdots  &   0 \\
-b_3  & b_2  & -b_1     & b_0 & \cdots  &   0 \\ 
-b_5  & b_4  & -b_3     & b_2 & \cdots  &   0 \\
-b_7  & b_6  & -b_5     &  b_4 & \cdots &    0 \\
\vdots & \vdots & \vdots  &  \vdots & \ddots & \vdots \\
-b_{2q-1} & b_{2q-2} & -b_{2q-3} &  b_{2q-4} & \cdots   & (-1)^qb_q
\end{array}
\right\rvert .
\label{eq:routh}
\end{align}
The first few Hurwitz determinants in terms of the principal minors $b_q$ are:
\begin{flalign}
\label{eq:routh1}   \Delta_1&= -b_1\\
\label{eq:routh2}   \Delta_2&= -b_1b_2+b_0b_3\\
\label{eq:routh3}   \Delta_3&= b_1b_2b_3-b_0b_3b_3+b_0b_1b_5-b_1b_1b_4\\
\label{eq:routh4}   \Delta_4&= b_1b_2b_3b_4-b_0b_3^2b_4-b_1^2b_4^2-b_1b_2^2b_5+b_0b_2b_3b_5\notag\\
& \hphantom{=}+2b_0b_1b_4b_5-b_0^2b_5^2+b_1^2b_2b_6-b_0b_1b_3b_6-b_0b_1b_2b_7\notag\\
& \hphantom{=}+b_0^2b_3b_7.
\end{flalign}
Upon use of the purely cycle-based expressions for the principal minors (Eq.~\ref{eq:minor}), we obtain for the first two determinants:
\begin{flalign}
\label{eq:routh1cycle}   \Delta_1&= -c_1\\
\label{eq:routh2cycle}   \Delta_2&= -c_1\cdot\overline{c_1c_1}+c_0\cdot\overline{c_1c_1c_1} \nonumber\\
&\hphantom{=}  +c_1\cdot c_2 -c_0\cdot\overline{c_1c_2} \nonumber\\
&\hphantom{=}  +c_0\cdot c_3.
\end{flalign}
For a network with $n\leq3$ species, the terms that potentially contribute to the third determinant (corresponding to the first two terms of Eq.~\ref{eq:routh3}) are:
\begin{flalign}
\Delta_3         &=    c_1\cdot\overline{c_1c_1}\cdot\overline{c_1c_1c_1}-c_0\cdot \overline{c_1c_1c_1}\cdot\overline{c_1c_1c_1}  \nonumber\\
     &\hphantom{=}  -c_1\cdot\overline{c_1c_1}\cdot\overline{c_1c_2} -c_1\cdot\overline{c_1c_1c_1}\cdot c_2 +2c_0\cdot\overline{c_1c_1c_1}\cdot\overline{c_1c_2}   \nonumber\\
     &\hphantom{=} +c_1\cdot c_2\cdot\overline{c_1c_2} - c_0\cdot \overline{c_1c_2}\cdot\overline{c_1c_2}        \nonumber\\
     &\hphantom{=} +c_1\cdot\overline{c_1c_1}\cdot c_3 -2c_0\cdot\overline{c_1c_1c_1}\cdot c_3 \nonumber\\
     &\hphantom{=} -c_1\cdot c_2\cdot c_3 + 2c_0\cdot \overline{c_1c_2}\cdot c_3\nonumber\\
     &\hphantom{=} -c_0\cdot c_3\cdot c_3.
\label{eq:routh3cycle}
\end{flalign}
In the above expressions, $c_0\equiv b_0$ (not to be confused with the less meaningful ``placeholder'' $c_0$ used to compute the cycle partitions in Eq.~\ref{eq:minor}). The raised dot indicates normal multiplication. Terms containing the same suite of cycles are shown in the same line to indicate where cancellations can occur. Finding an appropriate topological notation that allows removal of all such potential cancellations, thereby reducing these expressions even further, remains an open problem \cite{clarke_graph_1974,clarke_stability_1980}.

Orlando's formula \cite{orlando_sul_1911,gantmacher_applications_1959} can be used to express the penultimate and ultimate Hurwitz determinants for a network with $n$ species as:
\begin{align}
\label{eq:orlando1}   \Delta_{n-1} &= (-1)^{\frac{n(n-1)}{2}}a_0^{n-1}\prod_{1\leq i<k\leq n}(\lambda_i + \lambda_k)\\
\label{eq:orlando2}   \Delta_{n} &= (-1)^{\frac{n(n-1)}{2}}a_0^n\lambda_1\ldots\lambda_n\prod_{1\leq i<k\leq n}(\lambda_i + \lambda_k).
\end{align}
These formulas importantly indicate that $\Delta_{n-1}=\Delta_n=0$ upon appearance of a pair of purely complex roots, representing a necessary condition for a Poincar\'e-Andronov-Hopf bifurcation \cite{strogatz_nonlinear_1994} (hereafter, referred to simply as a Hopf bifurcation).

While the Li$\acute{\textrm{e}}$nard-Chipart conditions \cite{lienard_sur_1914,gantmacher_applications_1959,jury_j.j._1996} would be even simpler to verify, I will retain the full Hurwitz determinants due to their more straightforward application for root counting using the Routh array discussed below.

\section{Unstable root counting through use of the Routh array}
The number of unstable roots, $k$, with positive real part is equal to the number of sign changes in the first column of the Routh array \cite{gantmacher_applications_1959}, which in terms of the Hurwitz determinants is: 
\begin{equation}
k=V\left(a_0,\Delta_1,\frac{\Delta_2}{\Delta_1},\frac{\Delta_3}{\Delta_2},\ldots,\frac{\Delta_n}{\Delta_{n-1}}\right).
\label{eq:routharray}
\end{equation}
This criterion clearly fails, or is ambiguous, if any of the Hurwitz determinants equals 0. For these cases, the following generalization must be taken \cite{gantmacher_applications_1959}. Consider a string of $p$ Hurwitz determinants that are all zero. If this string terminates at $\Delta_n$, then one can truncate the Routh array, applying the above criterion for the determinants $\Delta_1,\ldots,\Delta_{n-p}$. If, however, $\Delta_n\neq0$ (and therefore $\Delta_{n-1}\neq0$, see Eqs.~\ref{eq:orlando1} and \ref{eq:orlando2}), with the string extending from $\Delta_{s+1},\ldots,\Delta_{s+p}$ ($\Delta_s\neq 0$ and $\Delta_{s+p+1}\neq0$), then
\begin{align}
k&=V\left(a_0,\Delta_1,\frac{\Delta_2}{\Delta_1},\ldots,\frac{\Delta_s}{\Delta_{s-1}}\right)\notag\\
&  \hspace{1.1em}+\frac{p+1}{2}+\frac{1}{2}\left(1-(-1)^{(p+1)/2}\mathrm{sign}\left(\frac{\Delta_s}{\Delta_{s-1}}\frac{\Delta_{s+p+2}}{\Delta_{s+p+1}}\right)\right)\notag\\
& \hspace{1.1em}+V\left(\frac{\Delta_{s+p+2}}{\Delta_{s+p+1}},\ldots,\frac{\Delta_n}{\Delta_{n-1}}\right).
\label{eq:routharraymod}
\end{align}
The second line of the above is equal to $(p+1)/2$ if $(-1)^{(p+1)/2}$ times the ``sign'' term yields $+1$, or $(p+3)/2$ if this product yields $-1$ ($p$ will always be odd). For $s=0$, the above formula should be modified to:
\begin{align}
k&=\frac{p+1}{2}+\frac{1}{2}\left(1-(-1)^{(p+1)/2}\mathrm{sign}\left(a_0\frac{\Delta_{p+2}}{\Delta_{p+1}}\right)\right)\notag\\
& \hspace{1.1em}+V\left(\frac{\Delta_{p+2}}{\Delta_{p+1}},\ldots,\frac{\Delta_n}{\Delta_{n-1}}\right).
\label{eq:routharraymods0}
\end{align}
In the below, the function $V$ will be expressed using only `$+$' or `$-$' signs as arguments \cite{gantmacher_applications_1959}, e.g. $V(+,-,+)$ (which has two sign changes and therefore two unstable eigenvalues). 

\section{Reactions}
Chemistry presents the interesting notion of a \textit{reaction}, for which I introduce the following mathematical generalization.
The $f_j(x_1,\ldots,x_n)$ in Eq.~\ref{eq:ODE} can be expressed as the sum over $m$ real reaction functions $v_k$ multiplied by species-specific stoichiometric scalars $s^k_j$. 
\begin{equation}
f_j(x_1,\ldots,x_n)=\sum_{k=1}^m v_k s^k_j,
\label{eq:ftov}
\end{equation}
with
\begin{equation}
H_{ij}=\left(\frac{\partial f_j}{\partial x_i}\right)_s= \sum_{k=1}^m \frac{\partial v_k}{\partial x_i}s^k_j.
\label{eq:ftovdiff}
\end{equation}
Both the reaction functions and the stoichiometric scalars can be positive, negative, or zero. The $v_k$ represent completely arbitrary real functions of a subset of the $x_i$. For the reaction derivatives $\partial v_k/\partial x_i$ in the above definition of $H_{ij}$, explicit reference to the particular steady state $s$ has been dropped for notational convenience (both here and in all subsequent expressions and figures). The intrinsic \textit{bipartite} nature of the reaction network topology has its fundamental basis in the separability of the $H_{ij}$ transition elements into distinct \textit{input} Jacobian terms ($i$) and \textit{output} stoichiometry terms ($j$) for each reaction $k$. 

Banaji \& Craciun \cite{banaji_graph-theoretic_2009,banaji_graph-theoretic_2010} show that the above result can be further generalized to the case in which the $f_j$ are completely arbitrary functions of the reactions $v_k$ (not simply a sum) through use of the chain rule, with 
\begin{equation}
s^k_j \equiv \frac{\partial f_j}{\partial v_k}
\end{equation}
in Eq.~\ref{eq:ftovdiff}.
While mathematically interesting, the practical value of such further generalization remains unclear and is not further explored in the current manuscript where I will assume throughout the above definition of each $f_j$ as a sum of reactions (Eq.~\ref{eq:ftov}).

\section{Principal minors as products of non-overlapping bipartite cycles}
Based on the preceding, the principal minors of $\mathbf{H}$ (Eq.~\ref{eq:Leibniz}) can be rewritten as:
\begin{align}
b_q &= \sum_{i_1<\cdots<i_q} \sum_{\pi(j_1,\ldots,j_q)} \epsilon_{j_1\ldots j_q}H_{i_1j_1}\cdots H_{i_qj_q}     \notag\\ 
       &= \sum_{i_1<\cdots<i_q} \sum_{\pi(j_1,\ldots,j_q)} \epsilon_{j_1\ldots j_q} \times       \notag\\
       & \hspace{2cm} \left(\sum_{k_1}\frac{\partial v_{k_1}}{\partial x_{i_1}}s^{k_1}_{j_1} \right)\cdots   \left(\sum_{k_q} \frac{\partial v_{k_q}}{\partial x_{i_q}}s^{k_q}_{j_q} \right)      \notag\\   
       &= \sum_{i_1<\cdots<i_q} \sum_{k_1,\ldots, k_q}  \frac{\partial v_{k_1}}{\partial x_{i_1}} \cdots \frac{\partial v_{k_q}}{\partial x_{i_q}} \times\notag\\
       &\hspace{2.3cm}\sum_{\pi(j_1,\ldots,j_q)} \epsilon_{j_1\ldots j_q}  s^{k_1}_{j_1} \cdots s^{k_q}_{j_q},
\label{eq:stoichiometry}
\end{align}
or, equivalently:
\begin{align}
b_q &= \sum_{j_1<\cdots<j_q} \sum_{\pi(i_1,\ldots,i_q)} \epsilon_{i_1\ldots i_q}H_{i_1j_1}\cdots H_{i_qj_q}    \notag                \\ 
       &= \sum_{j_1<\cdots<j_q} \sum_{k_1,\ldots, k_q}    s^{k_1}_{j_1} \cdots s^{k_q}_{j_q}   \times\notag\\
       & \hspace{1.7cm} \sum_{\pi(i_1,\ldots,i_q)} \epsilon_{i_1\ldots i_q}    \frac{\partial v_{k_1}}{\partial x_{i_1}} \cdots \frac{\partial v_{k_q}}{\partial x_{i_q}}. 
\label{eq:jacobian}
\end{align}
In both cases, it is clear that $k_1,\ldots,k_q$ should all be distinct, otherwise they will cancel with related terms under the permutation. This leads to the following important generalization of the ``non-overlapping'' aspect of the cycle products in Eq.~\ref{eq:minor} to \textit{bipartite} graphs: To avoid cancellations among the terms that comprise a particular principal minor, each cycle product should contain each species no more than once \textit{and} each reaction no more than once. This important result is equivalent to the topological specification of non-overlapping cycles in the full bipartite graph corresponding to the influence topology.

In Eq.~\ref{eq:stoichiometry}, specific stoichiometric subnetworks (defined by a particular species subset $i_1,\ldots,i_q$ and reaction subset $k_1,\ldots,k_q$) will not contribute if their determinant is zero. Similarly, in Eq.~\ref{eq:jacobian}, reaction derivative subnetworks will not contribute if their determinant is zero. A zero determinant indicates that the basis vectors of the subnetwork span a volume that is of dimension lower than $q$. Mass conservation in only a partial graph of the subnetwork is sufficient to generate a zero determinant for the stoichiometric terms \cite{schuster_what_1995,heinrich_regulation_1996}; however, there are many other ways that a zero determinant of the stoichiometry can be obtained. For example, for the network given by $\dot{x}_1=v_1$ and $\dot{x}_2=2v_1$, the quantity $2x_1-x_2$ will be conserved, not the total mass (e.g. $x_1+x_2$). A zero determinant for the Jacobian matrix is also possible, but in practice rarer to obtain and more difficult to immediately recognize based only on cursory inspection of the original governing equations, as the values of the Jacobian reaction derivatives (for nonlinear reactions anyway) will generally differ at each steady state, with the exact locations of the steady states therefore also necessary to know.

\section{Comparison of unipartite and bipartite graphs}
\begin{figure}[]
\vspace{0cm}
\centerline{     \includegraphics*[width=0.5\textwidth]{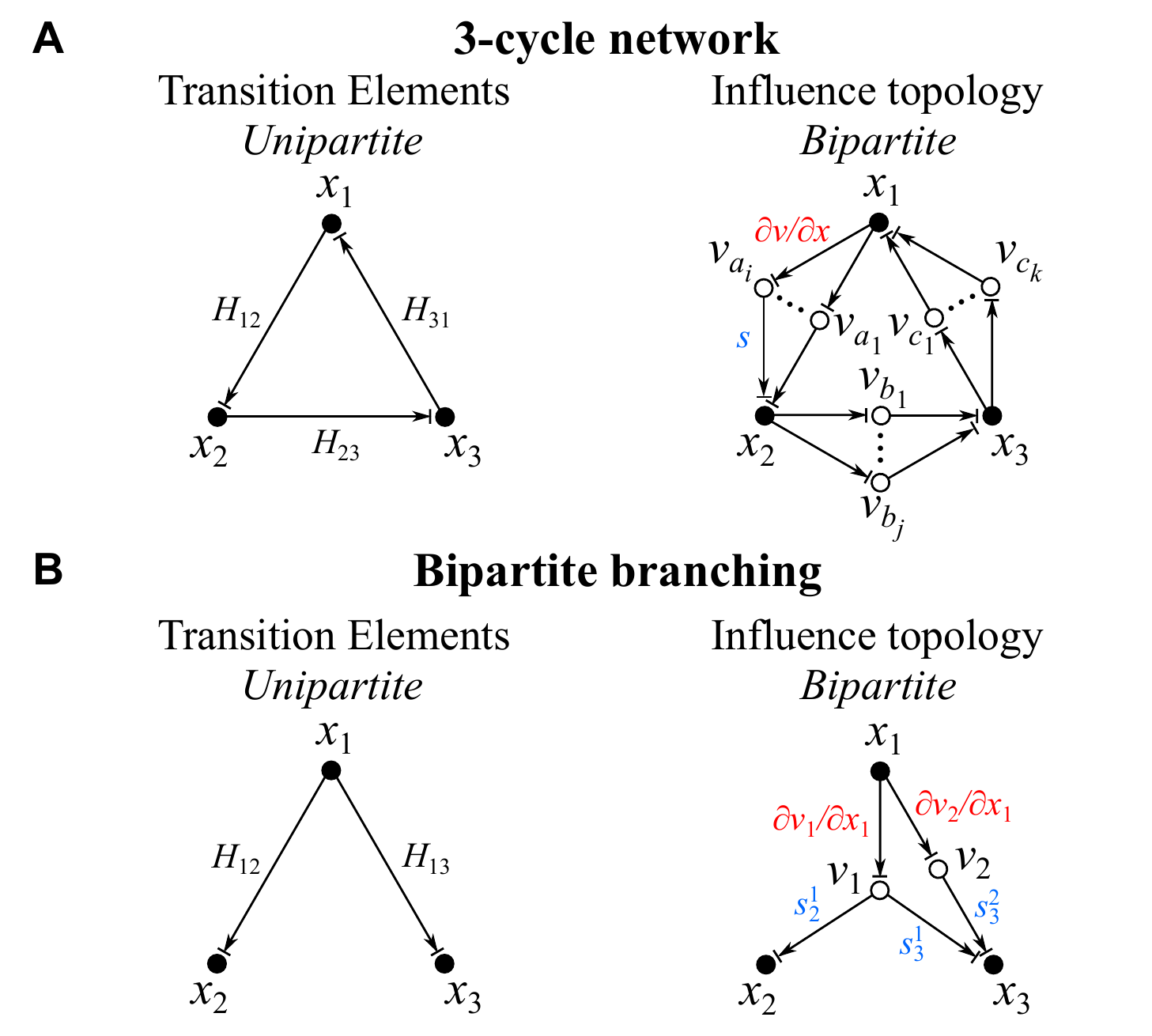}}
\vspace{-0.2cm}
      \caption{(A) Unipartite versus bipartite representations of a 3-cycle. The unipartite directed signed graph of the transition matrix elements of a network comprised of a single 3-cycle is shown, with the positive/negative influence from one species on another depicted graphically as the fusion of a normal arrow with a blunt arrow (graphical equivalent of `$\pm$'). The bipartite representation according to the influence topology is also displayed for a general 3-cycle network. In this representation, the directed edges from species to reaction nodes correspond to the Jacobian elements, $\partial v/\partial x$, and directed edges from the reactions to the species nodes correspond to the stoichiometries, $s$. (B) Unipartite versus bipartite depiction of a branching influence of one species on two immediately downstream species. In the unipartite representation, the separate influences of $x_1$ on $x_2$ and of $x_1$ on $x_3$ appear independent. In the bipartite representation, this apparent independence may no longer be true, as a single reaction can communicate a similar influence to both downstream species. The presence of $v_2$ in the displayed influence topology gives each downstream influence a degree of independence, but removal of $v_2$ is also a possible scenario, with both species then being regulated identically by the single reaction $v_1$.}
      \label{fig:bipartite}
\end{figure}

Consider, in isolation, a cycle of length $l$  that contributes to one of the cycle products in Eqs.~\ref{eq:Leibniz} and \ref{eq:minor} and that connects in an ordered fashion the species $x_{i_1} x_{i_2} x_{i_3} \ldots x_{i_l} x_{i_1}$: 
\begin{align}
c_l&=H_{i_1i_2}\cdots H_{i_l i_1}\notag\\
&=\left(\sum_{k_1}\frac{\partial v_{k_1}}{\partial x_{i_1}}s^{k_1}_{i_2} \right)  \cdots \left(\sum_{k_l} \frac{\partial v_{k_l}}{\partial x_{i_l}}s^{k_l}_{i_1} \right) \notag\\
&=\sum_{k_1,\ldots,k_l}\left(\frac{\partial v_{k_1}}{\partial x_{i_1}}s^{k_1}_{i_2}   \cdots  \frac{\partial v_{k_l}}{\partial x_{i_l}}s^{k_l}_{i_1} \right).
\end{align}
The last version consistitutes a sum over all unique cycles over the ordered species. This is made explicit for a network comprised of a single 3-cycle in Fig.~\ref{fig:bipartite}A, where the positive or negative effect of one species on another is illustrated by the superposition of an arrow (positive) and a blunt arrow (negative), the graphical equivalent of `$\pm$'. The 3-cycle defined by the unipartite product $H_{12}H_{23}H_{31}$ in Fig.~\ref{fig:bipartite}A is equivalent to the sum over all unique bipartite cycles. For example, the bipartite cycles $x_1 v_{a_1} x_2 v_{b_1} x_3 v_{c_1} x_1$ and $x_1 v_{a_2} x_2 v_{b_1} x_3 v_{c_1} x_1$ provide distinct contributions due to their paths through the different reactions $v_{a_1}$ and $v_{a_2}$. That the bipartite graph conveys more complete information about the network than the unipartite graph is illustrated in Fig.~\ref{fig:bipartite}B, in which the direct influence of one species on two other downstream species is depicted. For the unipartite graph, the branches to each individual species appear independent. For the bipartite version, a single reaction can affect both species (though other reactions can be species specific). As  reaction-like terms (and the parameters that define them) provide a useful descriptor of network dynamics in many different fields, the bipartite graph provides the most complete representation of how the network of reactions communicates the influence of one species on its immediately downstream species.

\section{Transition matrix as product of Jacobian and stiochiometry matrices} 
Through use of Eq.~\ref{eq:ftovdiff}, Eq.~\ref{eq:matrixform} can be reexpressed as:
\begin{align}
\lefteqn{\frac{d}{dt}(\Delta x_1,\ldots,\Delta x_n)\simeq(\Delta x_1,\ldots,\Delta x_n)\times}\notag\\
&&\left(\begin{array}{ccc}
\sum_k \frac{\partial v_k}{\partial x_1} s_1^k  & \cdots    & \sum_k \frac{\partial v_k}{\partial x_1} s_n^k \\
\vdots & \ddots & \vdots  \\
\sum_k \frac{\partial v_k}{\partial x_n} s_1^k     &  \cdots  & \sum_k \frac{\partial v_k}{\partial x_n} s_n^k 
\end{array} \right) \notag\\
\lefteqn{\frac{d}{dt}(\Delta x_1,\ldots,\Delta x_n)\simeq(\Delta x_1,\ldots,\Delta x_n)\times}\notag\\
&&\left(\begin{array}{ccc}
\frac{\partial v_1}{\partial x_1}  & \cdots    & \frac{\partial v_m}{\partial x_1}  \\
\vdots & \ddots & \vdots  \\
\frac{\partial v_1}{\partial x_n}      &  \cdots  &  \frac{\partial v_m}{\partial x_n}  
\end{array} \right)
\left(\begin{array}{ccc}
s^1_1  & \cdots    & s^1_n \\
\vdots & \ddots & \vdots  \\
s^m_1  &  \cdots  &  s^m_n  
\end{array} \right).
\label{eq:influencetopology}
\end{align}
This result can also be directly obtained from the definition of the reactions in Eq.~\ref{eq:ftov}, which can be recast in matrix form as a reaction row vector multiplied on the right by the stoichiometric matrix. Perturbation from the steady state can then be written as a species perturbation row vector times the reaction Jacobian matrix, $\mathbf{J}$ (derivatives of the reactions with respect to the species), times the stoichiometric matrix, $\mathbf{s}$, leading to the exact same expression as above. That $\mathbf{H}$ is obtained by multiplication of $\mathbf{J}$ and $\mathbf{s}$ implies by the Cauchy-Binet theorem that $\mathrm{rank}(\mathbf{H})\leq \min{\{\mathrm{rank}(\mathbf{J}),\mathrm{rank}(\mathbf{s})\}}$. A rank of $\mathbf{H}$ strictly less than the separate ranks of $\mathbf{J}$ and $\mathbf{s}$ can be obtained if the product of $\mathbf{J}$ with $\mathbf{s}$ leads to an additional loss of dimensionality.

\section{Stoichiometric scaling} Without loss of generality, the rows of the stoichiometry matrix in Eq.~\ref{eq:influencetopology} can be scaled, with a corresponding inverse scaling of the Jacobian matrix columns, to obtain:
\begin{align}
\lefteqn{\frac{d}{dt}(\Delta x_1,\ldots,\Delta x_n) \simeq(\Delta x_1,\ldots,\Delta x_n)\times} \nonumber\\
&&\left(\begin{array}{ccc}
\alpha_1\frac{\partial v_1}{\partial x_1}  & \cdots    & \alpha_m\frac{\partial v_m}{\partial x_1}  \\
\vdots & \ddots & \vdots  \\
\alpha_1\frac{\partial v_1}{\partial x_n}      &  \cdots  &  \alpha_m\frac{\partial v_m}{\partial x_n}  
\end{array} \right)
\left(\begin{array}{ccc}
\frac{s^1_1}{\alpha_1}   & \cdots    & \frac{s^1_n}{\alpha_1} \\
\vdots & \ddots & \vdots  \\
\frac{s^m_1}{\alpha_m}   &  \cdots  &   \frac{s^m_n}{\alpha_m}  
\end{array} \right)\nonumber\\
\lefteqn{\frac{d}{dt}(\Delta x_1,\ldots,\Delta x_n) \simeq(\Delta x_1,\ldots,\Delta x_n)\times} \nonumber\\
&&\left(\begin{array}{ccc}
r^1_1  & \cdots    & r^m_1  \\
\vdots & \ddots & \vdots  \\
r^1_n      &  \cdots  & r^m_n 
\end{array} \right)
\left(\begin{array}{ccc}
\pm\sigma^1_1   & \cdots    & \pm\sigma^1_n \\
\vdots & \ddots & \vdots  \\
\pm\sigma^m_1  &  \cdots  &   \pm\sigma^m_n
\end{array} \right),
\end{align}
with
\begin{align}
\alpha_k &= \left|s^k_{j_k}\right|\\
\sigma^k_j &= \frac{1}{\alpha_k}\left|s^k_j\right|\\
r^k_i &= \alpha_k \frac{\partial v_k}{\partial x_i}.
\end{align}
In the above, the $j_k$ refer to a particular non-zero stoichiometry of reaction $k$.  Scaling of the stoichiometry matrix merely amounts to a redefinition of the reactions such that at least one of the scaled stoichiometries of each reaction is $\pm1$. For typical networks of interest, the $\boldsymbol{\sigma}$ matrix will be sparsely filled with elements that are either simply $\pm1$ or $\pm1$ times a finite number of non-negative scale factors $\sigma_1$,  \ldots, $\sigma_g$. Similarly, the $\boldsymbol{r}$ matrix will be sparsely filled with elements equal to the real variables $r_1$, \ldots, $r_f$ ($r_0$ may also appear; see the discussion below regarding cycle compaction). Depending on the sign of the monotonicity of the reaction with respect to the particular species, the $r_i$ can be negative or positive. If the reaction has a universal monotonicity over the entire phase space, then we can fix this edge with an arrow (always positive monotonicity) or blunt arrow (always negative) corresponding to a single fixed influence topology (Fig.~\ref{fig:topologies}). If the monotonicity is not universal, this uncertainty in sign will be conveyed through the superposition of an arrow and a blunt arrow as in Fig.~\ref{fig:bipartite}.

\begin{figure}[]
\vspace{0cm}
\centerline{     \includegraphics*[width=0.5\textwidth]{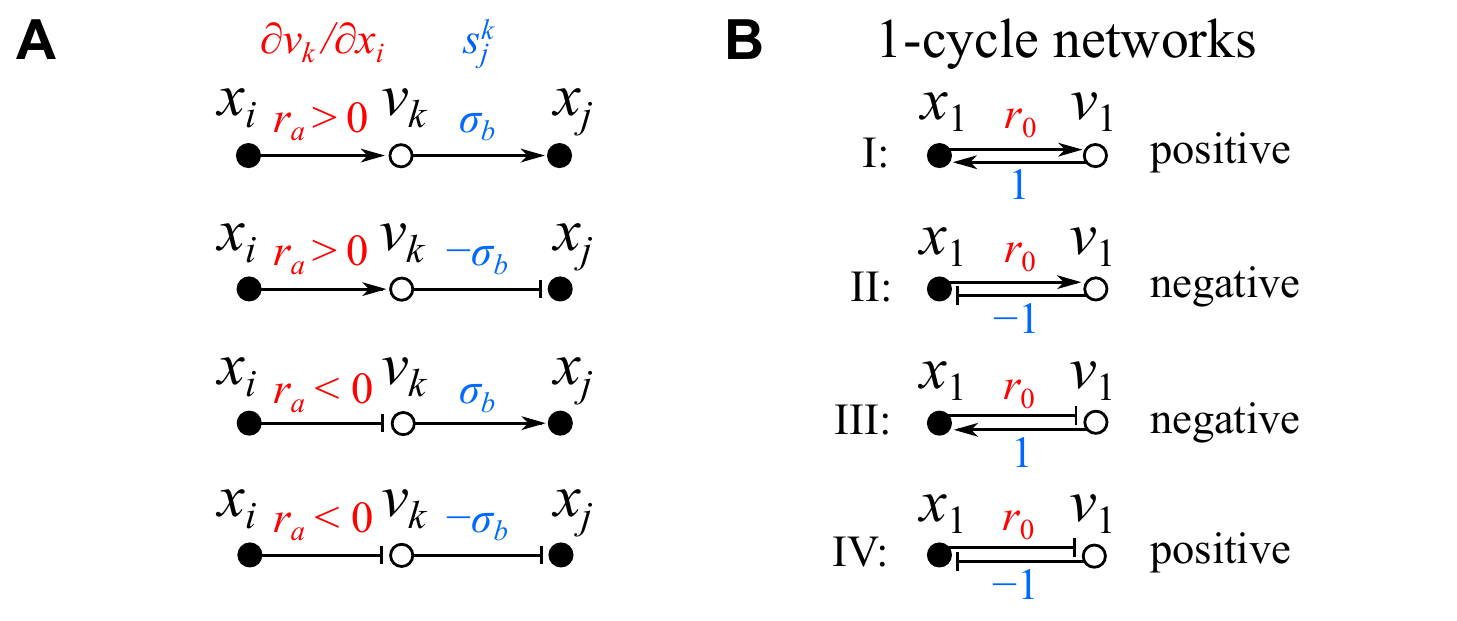}}
\vspace{-0.2cm}
      \caption{(A) All possible signed directed bipartite connections between two species. (B) All possible 1-cycle networks.}
      \label{fig:1cycles}
\end{figure}

The complete set of possible connections between two species using the above parameters is given in graphical terms in Fig.~\ref{fig:1cycles}A; these basic graphical elements underlie the mathematically rigorous definition of a network's \textit{influence topology}.

\section{Cycle compaction}
\begin{figure}[]
\vspace{0cm}
\centerline{\includegraphics*[width=0.5\textwidth]{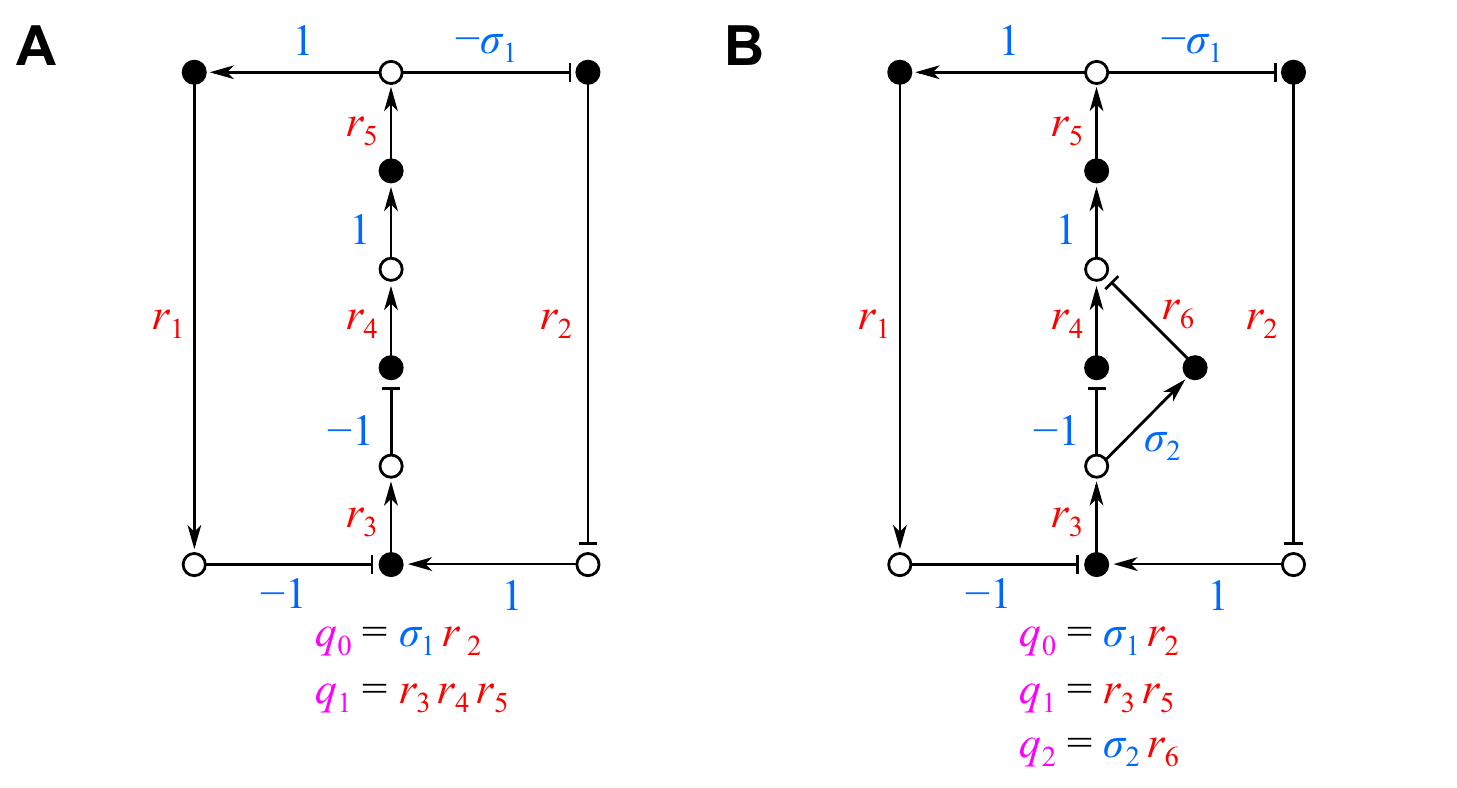}}
\vspace{-0.2cm}
      \caption{Cycle compaction. (A) For a network composed of two overlapping cycles, the cycle compaction terms ($q_0$ and $q_1$) are listed. (B) Upon slight modification of A, a network composed of four unique overlapping cycles is obtained, leading to the splitting up of the cycle compaction term $q_1$ in A into the new $q_1$ and $q_2$ terms listed here. Interestingly, the new $q_1$ is the product of the widely separated graphical elements $r_3$ and $r_5$, emphasizing the fact that the overlapping region between two (or more) cycles, or the non-overlapping region of a single cycle (see the Jenkin-Maxwell network discussed below), might not be contiguous.}
      \label{fig:cycle_compaction}
\end{figure}

The individual $r_i$ and $\sigma_j$ terms appear in the Routh-Hurwitz conditions only through their contributions to complete cycles in the graph. Each cycle is simply a number representing the product of individual Jacobian edges ($r_i$) and stoichiometry edges ($\pm1$ or $\pm\sigma_j$). Consider an arbitrary graph with $n$ species, $m$ reactions, $J$ Jacobian edges, and $S$ stoichiometric edges. Stoichiometric scaling reduces the total number of parameters needed to describe the influence topology to $J+S-m$. As cycles are the important objects for determining network stability (see Eq.~\ref{eq:minor}), not the individual edge parameters, additional dimensional reduction can often be achieved through \textit{cycle compaction}, which consists of the expression of a product of multiple $r_i$ and/or $\sigma_j$ terms as a single cycle compaction term $q_k$. Such cycle compaction can be easily understood from either an algebraic or a topological perspective. From an algebraic perspective, write down all cycles in the graph in terms of their products of $r_i$ and/or $\sigma_j$ terms. Certain combinations of these terms may always appear together, allowing their replacement by a single $q_k$ variable ($q_1$, $q_2$, \ldots). From a topological perspective, consider each non-overlapping and overlapping part of all cycles. For a given non-overlapping part of a cycle (whether contiguous or non-contiguous), if more than one $r_i$ and/or $\sigma_j$ appears, their product can be replaced with a $q_k$ term. For a given overlap between two (or more) cycles, find the largest overlapping region (again, contiguous or non-contiguous) that is completely shared by \textit{all} overlapping cycles and replace the product of the $r_i$ and/or $\sigma_j$ terms that define this region with a $q_k$ term. These simple and intuitive topological rules are illustrated in Fig.~\ref{fig:cycle_compaction}.

\section{Temporal scaling} A final degree of freedom can be removed by temporal scaling, $\tau=\beta t$, with $\beta=|r_0|$ (scaling to a particular Jacobian element $r_0$) or $\beta= |q_0|^{1/z}$ (scaling to the particular cycle compaction term $q_0$ containing $z\geq1$ Jacobian elements):
\begin{align}
\lefteqn{\hspace{-0.5cm}\frac{d}{d\tau}(\Delta x_1,\ldots,\Delta x_n)\simeq(\Delta  x_1,\ldots, \Delta x_n)\times} \nonumber\\
&&\left(\begin{array}{ccc}
\rho^1_1  & \cdots    & \rho^m_1  \\
\vdots & \ddots & \vdots  \\
\rho^1_n      &  \cdots  & \rho^m_n 
\end{array} \right)
\left(\begin{array}{ccc}
\pm\sigma^1_1   & \cdots    & \pm\sigma^1_n \\
\vdots & \ddots & \vdots  \\
\pm\sigma^m_1  &  \cdots  &   \pm\sigma^m_n
\end{array} \right),
\end{align}
with $\rho_i^k=r_i^k/\beta$. The subscript 0 on $r_0$ or $q_0$ will be used in all of the below graphs of the influence topology to indicate which term is used for temporal scaling (it will contribute $\pm1$ in any further calculations). This term (either $r_0$ or $q_0$) will be retained in the graph for two reasons: (1) to stress the arbitrary nature of this choice, and (2) to allow the convenient swapping of subscripts in the graph to find the most optimal choice (in concert with stoichiometric scaling and topological compaction) that generates the least number of remaining dimensions. The $\boldsymbol{\rho}$ matrix is sparsely filled with at most one $\pm1$ term and the real variables $\rho_1$, \ldots, $\rho_f$. Each $\rho_i$ should be considered strictly positive (arrow) or strictly negative (blunt arrow) for the definition of a single influence topology. Alternatively, specific $\rho_i$ may be considered to have uncertain sign (arrow/blunt arrow in the influence topology), implying a set of possible influence topologies.

\section{Orphan species} ``Orphan species'' (aka, \textit{clamped} species) are those that do not have any reactions as parents, i.e. they do not lie downstream of any reaction node in the influence topology. They often appear explicitly in chemical reaction networks as species that are buffered by an infinite bath. From a mathematical perspective, however, such species, if included in the network's influence topology graph, would only obscure the topological interpretation of the interaction network. Clamped species have the same mathematical status as a change in the coefficients governing the description of the reactions. For this reason, orphan species will not be displayed as explicit nodes in the influence topology.

\section{Childless species} ``Childless species'' are those that are not the parent of any reaction, i.e. they do not lie upstream of any reaction node in the influence topology. Such species often make an important physical contribution to many chemical networks through their contribution to total mass conservation. However, aside from this ``bookkeeping'' value, they play no important role in the mathematical description of the underlying network dynamics. Childless species can therefore be neglected even in the definition of the original network: The species $j$ that comprise the relevant set of ODE's in Eq.~\ref{eq:ODE} should be restricted to only those that affect other species in the network.

\section{Orphan reactions} ``Orphan reactions'' are reactions that have no species as their parents, i.e. they do not lie downstream of any species node in the influence topology. They therefore have no functional dependence on any of the species. Orphan reactions are mathematically equivalent to the addition of a (possibly different) constant value term to each of the equations governing the network (see Eqs.~\ref{eq:ODE} and \ref{eq:ftov}). Orphan reactions have no effect on the influence topology, as they are removed upon taking the first derivatives of the $f_j(x_1,\ldots,x_n)$ to describe the first-order perturbation of the network. They will be indicated in the graphs of the networks considered below by a single $V^0$ node with one or more dashed lines (having possibly different stoichiometries) connected to the relevant species. While these orphan reactions have no effect on the influence topology, they can nevertheless shift the location of the steady states in the species phase space and the dimensionally reduced stability phase space (introduced below) that corresponds to the influence topology. Addition of such constant terms to one or more of the species of a network is the simplest method for generating new dynamical behavior in a manner that preserves the network's influence topology \cite{angeli_combinatorial_2013}.

\section{Childless reactions} ``Childless reactions'' are reactions that have no species as children in the interconnected network under consideration (they appear in the list of all possible directed bipartite graphs). Such reactions, while possibly controlled by a subset of the species of the network under consideration (with which they would share a Jacobian edge), can have no effect on its dynamics and can therefore be removed from the graph of the network's influence topology. They can nevertheless affect other purely downstream species that are not under consideration. A biological example suffices to clarify this notion. Consider a transcriptional network comprised of a set of proteins that collectively regulate their own transcription (e.g., proteins that control the cell cycle). Such proteins can also control the transcription of several purely \textit{downstream} genes. The protein products of these downstream genes are assumed to have no effect on the dynamics of the transcription network; these downstream products are therefore childless species and can be removed from consideration. The transcriptional reactions that are solely responsible for producing these now removed childless species themselves become childless reactions and can also be removed.

\section{Influence topology separability} There are two major aspects of the separability of a given network's influence topology graph into distinct subgraphs. The first is trivial: If the influence topology consists of separable influence topology subgraphs, which are not linked to each other through any edges, then the dynamics of each minimal subgraph (at least at the level of the influence topology) can be examined independently. The second aspect involves the notion of upstream versus downstream. Consider a particular influence topology graph that can be partitioned into an upstream subgraph, which consists of purely interconnected cycles on one level, and a downstream subgraph, which may consist of multiple levels and for which all connections between the upstream and downstream subgraphs are through directed edges emanating from the upstream graph. In this case, only the upstream subgraph is fundamental. For example, if the upstream subgraph can oscillate autonomously, then oscillations in the downstream subgraph may merely reflect its entrainment rather than its own inherent topological properties. For this reason, it makes sense to restrict our consideration to only influence topologies for which no upstream/downstream partitioning is possible. This consideration, coupled with the fact that all sources (orphan species/reactions) and sinks (childless species/reactions) can be removed from consideration by the abovementioned rules, allow definition of the \textit{fundamental} set of influence topologies to consist of only interconnected cycles (all species/reaction nodes can be reached by directed travel from all other nodes). However, it should be reiterated here that the influence topology, while imposing important constraints on a network's dynamics, is of course not the full story. The exact location of the steady state(s) within the stability phase space (see below) defined by the influence topology requires consideration of the full functional form of a given network's reactions (not just their first derivatives) as well as inclusion of all orphan reactions (which were removed upon taking the reaction derivatives to generate the influence topology).

\section{Stability phase space dimensionality} The first-order stability of a network at a particular steady state is completely specified by the $d=S+J$ parameters that respectively define its total number of stoichiometric edges, $S$, and Jacobian edges, $J$, as well as the exact position of the steady state solution in the stability phase space. The stability phase space is defined over the dimensions remaining after dimensional reduction by stoichiometric scaling, cycle compaction, and temporal scaling. As already discussed above, stoichiometric scaling allows one stoichiometric factor for each reaction to be set to $\pm1$, with the others labeled as $\sigma_1,\ldots,\sigma_{S-m}$. After temporal scaling, normalization to either $r_0$ or the cycle compaction term $q_0$ (containing at least one Jacobian element) will scale the normalized parameter to $\pm1$. If $c$ dimensions can be reduced through cycle compactions, the final dimensionality will then be $d = (S-m) + (J-1) - c$. It is worth emphasizing that all of the above dimensional reductions were achieved purely through examination of the network's influence topology.

Other potentially useful forms of dimensional reduction are possible through symmetries in the influence topology graph. Exchanges of certain of the $r_i$ or $\sigma_j$ can be topologically shown to generate an identical set of non-overlapping cycle products that define the principal minors (\textit{topological symmetry}), allowing redefinition of multiple parameters into a single parameter, $\Psi$, which is symmetric with respect to the $r_i$ and/or $\sigma_j$ parameters that define it. As well, the complex algebraic structure of the Hurwitz determinant inequalities can allow for the algebraic gathering of multiple parameters into one parameter (\textit{Hurwitz reduction}), often allowing expression of the stability of an entire network in terms of an inequality involving a single parameter, $\Gamma$. It is possible that these Hurwitz reductions also have a topological explanation, but not one as simple as all of the others described above. Examples of networks containing topological symmetries or the possibility for Hurwitz reductions will be encountered below.

\section{1-cycle networks}

We first examine the simplest possible networks constructed from a single 1-cycle. The four possible 1-cycles are:
\begin{align}
\mathrm{I}: \hspace{0.3cm}  \dot{x}_1  &=   \pm V^0 + V^1_1\\
\mathrm{II}: \hspace{0.3cm}  \dot{x}_1  &=   \pm V^0 - V^1_1\\
\mathrm{III}: \hspace{0.3cm}  \dot{x}_1 & =   \pm V^0 + V^1_{\overline{1}}\\
\mathrm{IV}: \hspace{0.3cm}  \dot{x}_1 & = \pm V^0 - V^1_{\overline{1}},
\label{eq:1cycle}
\end{align}
with their dimensionally reduced topological representations given in Fig.~\ref{fig:1cycles}B (stoichiometric scaling leads to a $\pm1$ stoichiometric edge; temporal scaling by $|r_0|$ implies a $\pm1$ Jacobian edge). In the above, I introduce the following useful shorthand notation $V^k_{i_1\ldots i_h}\equiv v_k(x_{i_1},\ldots,x_{i_h})$. The subscripts in $V^k_{i_1\ldots\i_h}$ indicate a monotically increasing (normal subscript) or decreasing (overlined subscript) dependence on the $h$ different species that control the reaction; an underlined subscript will be used to indicate an uncertain sign of the monotonicity. Due to the $n=1$ dimensionality, only the first Routh-Hurwitz condition, $\Delta_1=-b_1=-c_1$, is necessary to consider for the above 1-cycle networks, giving for networks I-IV, respectively, $b_1=1,-1,-1,1$ and $\Delta_1=-1,1,1,-1$. Networks I and IV are therefore unstable and networks II and III are stable. Note that only the signs of the reaction stoichiometry and its monotonicity with respect to $x_1$ are necessary to specify to ascertain the network's stability. Addition of the static terms $\pm V^0$ can shift the steady-state solution but cannot otherwise affect the dynamics. For a reaction function $V^1_{\underline{1}}=v_1(x_1)$ that does not have a universal monotonicity (either increasing or decreasing with respect to $x_1$), the $x_1$ phase space can be partitioned into regions over which either $V^1_1$ or $V^1_{\overline{1}}$ holds, corresponding to a single influence topology in each region (this can of course be generalized to higher dimensional phase spaces as well). 

A simple explicit example of networks I--IV is given below:
\begin{align}
\mathrm{I}:  \hspace{0.3cm}    \dot{x}_1 & =  -1+x_1\\
\mathrm{II}:   \hspace{0.3cm}   \dot{x}_1 & =  \hphantom{+}1-x_1\\
\mathrm{III}:   \hspace{0.3cm}   \dot{x}_1 & =  -1+1/x_1\\
\mathrm{IV}:   \hspace{0.3cm}   \dot{x}_1 & =   \hphantom{+}1-1/x_1.
\end{align}
For all of these networks, additional constant terms have been added to position the single steady state solution at the positive value of $x_1^s=1$. Networks I and II correspond to the familiar examples of exponential growth and decay, respectively. Networks III and IV are perhaps more exotic, but, from the perspective of the influence topology, are equally fundamental. For these examples, I have chosen the particularly simple reaction functions proportional to $x_1$ and $1/x_1$, but any functions having the same stoichiometric sign and reaction monotonicity will have the same stability/instability (e.g. one could replace $x_1$ with $e^{x_1}$ in network I or $1/x_1$ with $1/\arctan{x_1}$ in network III).

\section{2-cycle networks}
\begin{figure}[]
\vspace{0cm}
\centerline{\includegraphics*[width=0.5\textwidth]{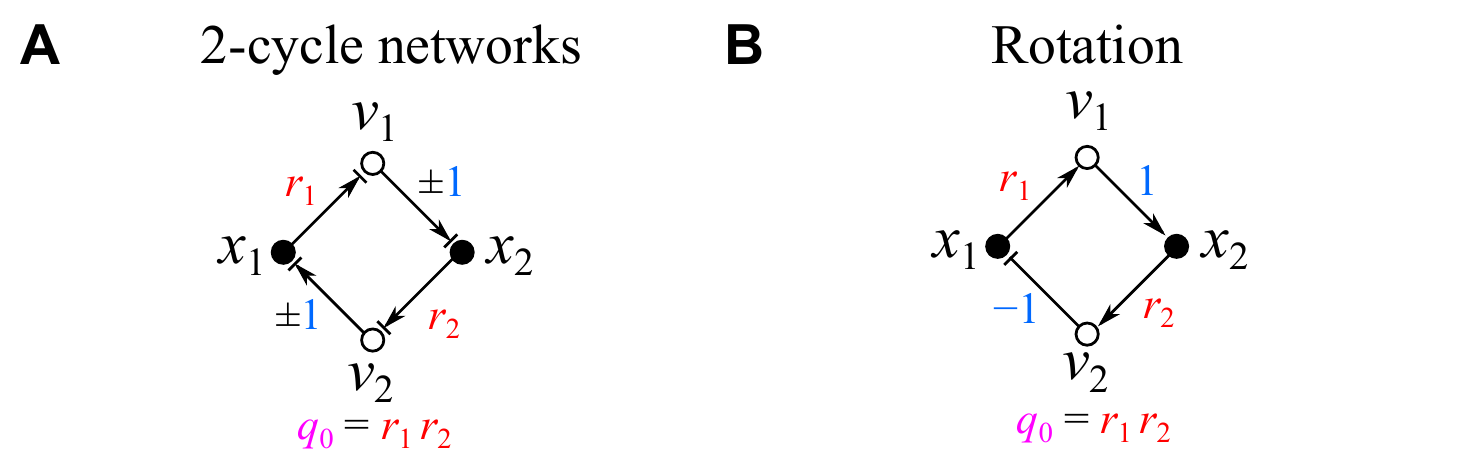}}
\vspace{-0.2cm}
      \caption{2-cycle networks. (A) All possible 2-cycle network influence topologies. (B) Rotation network influence topology.}
      \label{fig:2cycles}
\end{figure}

All possible 2-cycle networks are schematically represented in the single graph shown in Fig.~\ref{fig:2cycles} (the degeneracy of these networks will be addressed further below).  After cycle compaction (defining $q_0=r_1r_2$) and temporal scaling ($\rho_1=r_1/\sqrt{|q_0|}$ and $\rho_2=r_2/\sqrt{|q_0|}$), it is clear that $b_1=0$ and $b_2=-c_2=\mp1$. 
For all of these topologies, it is obvious that $\Delta_1=-b_1=0$ and $\Delta_2=-b_1b_2+b_0b_3=0$ due to the absence of 1-cycles in the network. That all Routh-Hurwitz conditions are equal to 0 implies that no information can be obtained from first-order perturbations about the steady state; higher order perturbations must be assessed to establish the stability of a given steady state.

An important example of a 2-cycle network is:
\begin{align}
\dot{x}_1   = && -k_1x_2      &&                      \label{eq:rotation1}\\
\dot{x}_2   = &&                    &&    k_2 x_1,    \label{eq:rotation2}
\end{align}
which, for $k_1=k_2$, corresponds to constant rotational motion at a fixed radius determined by the initial values (boundary conditions). 
We can rewrite the rotation network in a more general way as:
\begin{align}
\dot{x}_1   = &&  -V^1_2      &&                        \\
\dot{x}_2   = &&                    &&    V^2_1.
\end{align}
In the above, I again employ the shorthand notation for the reaction functions explained above, with $V^k_i$ corresponding to reaction $k$ with positive monotonic dependence on species $i$. The principal minors for this generalized network are $b_1=0$ and $b_2=-c_2=1$, which, as for the general case, leads to $\Delta_1=0$ and $\Delta_2=0$ and no information about steady state stability obtainable at first order. For the original rotation network (Eqs.~\ref{eq:rotation1} and \ref{eq:rotation2}), the linearity of the reactions implies that all higher order perturbations are trivially 0. The different solutions of this network depend on the initial conditions and foliate the $x_1$-$x_2$ phase space as circles of each possible radius centered on the origin. Inclusion of non-zero constant terms ($V^0$ terms) would merely shift the origin of these foliated circular trajectories. 

\section{$n$-cycle networks}
According to Eq.~\ref{eq:minor}, a network comprised of a single $n$-cycle will yield $b_n=c_n$ for odd $n$ and $b_n=-c_n$ for even $n$ with all other principal minors equal to 0. For even $n$, examination of the non-zero terms in the columns of the Routh-Hurwitz matrix (Eq.~\ref{eq:routh}) shows that columns containing $b_0=1$ and $-b_n$ alternate with all-zero columns, allowing the immediate conclusion that all of the Hurwitz determinants equal $0$, with no further information possible at first-order (the generalization of the above result obtained for 2-cycles). This result can actually be generalized further: For an influence topology comprised of only even cycles, all Hurwitz determinants are zero, implying no information is obtainable at first order. For odd $n\geq3$ cycle networks, column swapping of the Hurwitz determinants (Eq.~\ref{eq:routh}) to place the single non-zero term in each column (either $b_0=1$ or $-b_n$) along the diagonal can be shown to lead to the following general result:
\begin{align}
\Delta_1 &= 0\nonumber\\
     &\hspace{0.17cm}\vdots  \nonumber\\
\Delta_{n-2} & =  0\nonumber\\
\Delta_{n-1} &= (-1)^{\frac{n+1}{2}}c_n^{\frac{n-1}{2}}  \\
\Delta_n &= -(-1)^{\frac{n+1}{2}}c_n^{\frac{n+1}{2}}.
\end{align}
After cycle compaction and temporal scaling, $c_n=\pm1$. For $n=3$, $\Delta_2=\pm1$ and $\Delta_3=-1$; for $n=5$, $\Delta_4=-1$ and $\Delta_5=\pm1$. For $n=7$, this pattern repeats with $\Delta_6=\pm1$ and $\Delta_7=-1$.
From the Routh array (Eq.~\ref{eq:routharraymods0}), it can easily be shown that the number of unstable roots for an odd $n$-cycle network is:
\begin{equation}
k = \frac{1}{2}\left(n+(-1)^{(n-1)/2}c_n\right).
\end{equation}
For $n=3,7,11,\ldots$, this implies $k=(n-1)/2$ for $c_n=1$ and $k=(n+1)/2$ for $c_n=-1$. Oppositely, for $n=5,9,13,\ldots$, the above implies $k=(n-1)/2$ for $c_n=-1$ and $k=(n+1)/2$ for $c_n=1$.

\section{Analysis of classical networks} In the following, I provide detailed examinations of six classical networks from the diverse fields of control theory (Jenkin-Maxwell \cite{maxwell_governors_1868}), electronics (van der Pol \cite{van_der_pol_lxxxviii._1926}), ecology (Lotka-Volterra  \cite{lotka_analytical_1920,volterra_variazioni_1926,volterra_variations_1928}), chemistry (Brusselator \cite{prigogine_symmetry_1968}), biochemistry (Sel'kov \cite{selkov_self_1968}), and synthetic biology (Repressilator \cite{goodwin_oscillatory_1965,banks_stability_1978,smith_oscillations_1987,elowitz_synthetic_2000}). Each network is generalized to its dimensionally reduced influence topology, with its full stability phase space examined for regions in which one or more Hurwitz determinants are negative. Particular attention is paid to those regions in which both Hurwitz determinants $\Delta_{n-1}$ and $\Delta_{n}$ simultaneously go negative, a necessary condition for the presence of a Hopf bifurcation (representing a stronger and more general criterion than recent results on Hopf bifurcation exclusion obtained for DSR graphs by Angeli et al. \cite{angeli_combinatorial_2013}). For most of the networks, an explicit expression of the steady state solution for the original governing equations allows display of the actually accessible regions of the stability phase space. These detailed examinations raise several further issues discussed in greater detail below.

\section{Jenkin-Maxwell network} 
\begin{figure}[]
\vspace{-4.2cm}
\centerline{     \includegraphics*[width=0.5\textwidth]{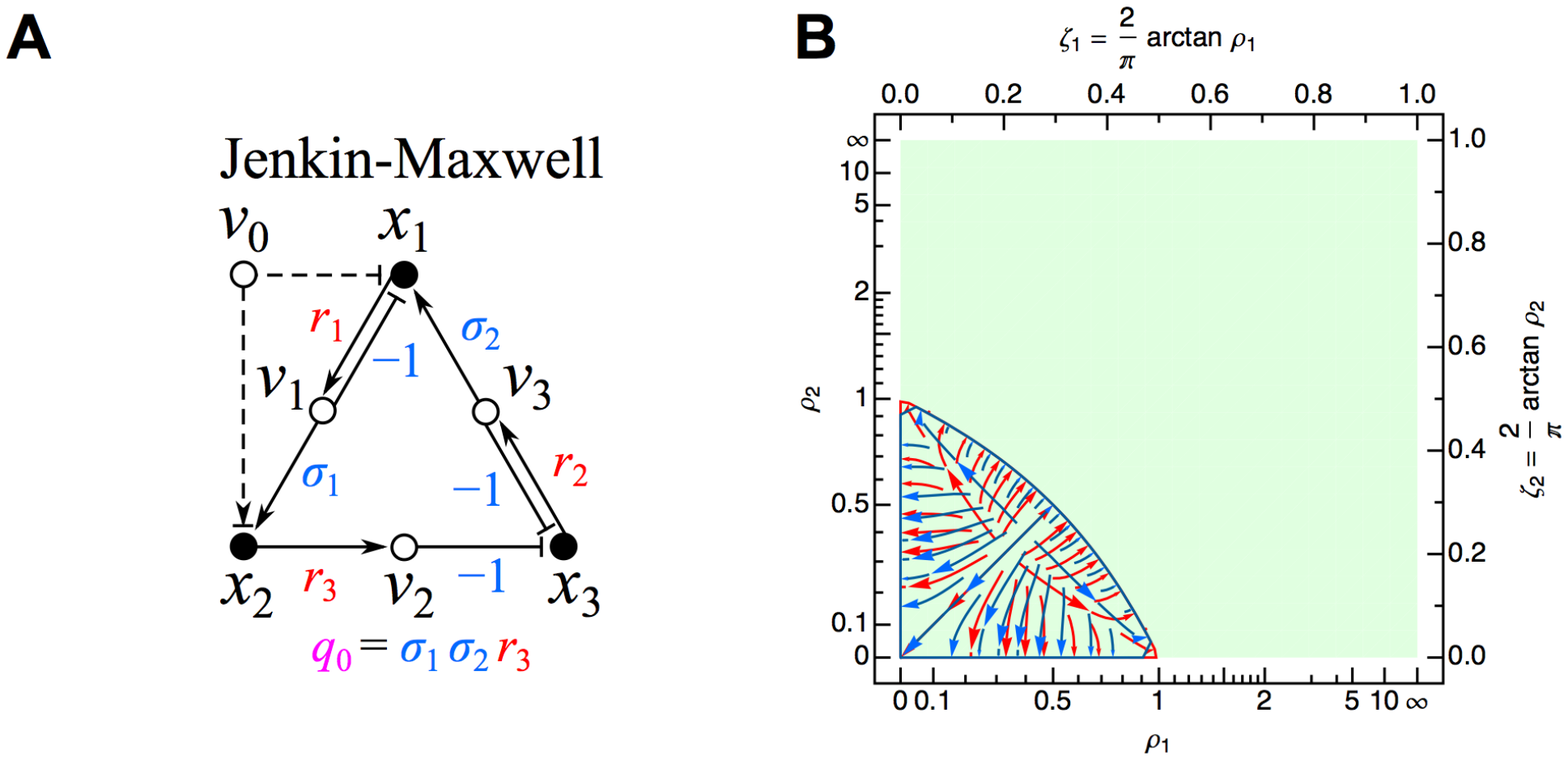}}
\vspace{-4.5cm}
      \caption{Jenkin-Maxwell network. (A) Influence topology. Cycle compaction allows definition of $q_0=\sigma_1\sigma_2 r_3$. Temporal scaling of all Jacobian edges to $|q_0|$ leaves only $\rho_1=r_1/|q_0|$ and $\rho_2=r_2/|q_0|$. (B) Stability phase space. Axes correspond to the two parameters, $\rho_1$ and $\rho_2$, that remain after dimensional reduction. For plotting $\rho_1$ and $\rho_2$, I have used the variable transformation $\zeta_i=\frac{2}{\pi}\arctan\rho_i$ to allow visualization of the entire range of the $\rho_i$ from 0 to $\infty$ (this arctan transform also automatically permits visualization of the range $-\infty<\rho_i<0$, which would correspond to a different sign for this Jacobian element and therefore a different influence topology). Flows in the plot map the zones over which $\Delta_1$ (black), $\Delta_2$ (red), and $\Delta_3$ (blue) are negative. Only $\Delta_2$ (red) and $\Delta_3$ (blue) can go negative (in this case, simultaneously). The green background indicates that $\rho_1$ and $\rho_2$ can independently assume any positive definite values based on their definitions in terms of the parameters used to define the original Jenkin-Maxwell equations (Eqs.~\ref{eq:jenkin-maxwell1}--\ref{eq:jenkin-maxwell3}).}
      \label{fig:jenkin-maxwell}
\end{figure}

In Maxwell's foundational paper on control theory from 1868 entitled ``On Governors'' \cite{maxwell_governors_1868}, he considered several examples of physical devices that worked to \textit{govern}, and importantly sustain, the angular velocity of a core component. For one such physical device described by Jenkin, Maxwell derived the following second order differential equations:
\begin{align}
\label{eq:jenkin-maxwellorig1}   B\frac{d^2y}{dt^2} &= F\left(\frac{dx}{dt}-V_1\right)-Y\frac{dy}{dt}-W\\
\label{eq:jenkin-maxwellorig2}   M\frac{d^2x}{dt^2} &= P-R-F\left(\frac{dx}{dt}-V_1\right)-Gy.
\end{align}
In the above, the nine parameters $B$, $F$, $V_1$, $Y$, $W$, $M$, $P$, $R$, and $G$ are all positive definite. Taking $x_1=dy/dt$, $x_2=y$, and $x_3=dx/dt$, these two second order equations reduce to the following three first-order equations:
\begin{align}
\label{eq:jenkin-maxwell1}  \dot{x}_1  = && -k_0                          &&   -k_1 x_1                  && +\sigma_2 k_2x_3       &&              \\
\label{eq:jenkin-maxwell2}  \dot{x}_2  = &&                                  &&    \sigma_1 k_1x_1    &&                                      &&                \\
\label{eq:jenkin-maxwell3}  \dot{x}_3  = && \pm \sigma_3k_0      &&                                   &&  -k_2x_3                       &&     -k_3x_2,
\end{align}
with $k_0=(FV_1+W)/B$, $k_1=Y/B$, $\sigma_2=M/B$, $\pm\sigma_3k_0=(FV_1+P-R)/M$, $k_2=F/M$, and $k_3=G/M$.
The generalized Jenkin-Maxwell network is:
\begin{align}
\dot{x}_1  = && -V^0                         &&   -V^1_1                    && +\sigma_2 V^2_3       &&              \\
\dot{x}_2  = &&                                 &&    \sigma_1 V^1_1    &&                                    &&                \\
\dot{x}_3  = && \pm \sigma_3 V^0    &&                                  &&  -V^2_3                      &&     -V^3_2,
\end{align}
corresponding to the influence topology shown in Fig.~\ref{fig:jenkin-maxwell}A, with principal minors: 
\begin{align}
b_1 &= c_1                                   \hspace{-1.8cm}    &&=-\rho_1-\rho_2    \\
b_2 &= \overline{c_1c_1}-c_2       \hspace{-1.8cm}  &&= \rho_1\rho_2  \\
b_3 &= \overline{c_1c_1c_1}-\overline{c_1c_2}+c_3     \hspace{-1.8cm}    &&= -\rho_1\rho_2.
\end{align}
In the above, $\rho_1=r_1/|q_0|$ and $\rho_2=r_2/|q_0|$ with $q_0=\sigma_1\sigma_2r_3$.
The principal minor $b_1$ is given by the sum of the two 1-cycles: $(r_1/|q_0|)(-1)=-\rho_1$ and $(r_2/|q_0|)(-1)=-\rho_2$. For $b_2$, only the first term (corresponding to two non-overlapping 1-cycles) contributes due to the absence of a 2-cycle in the network. For $b_3$, only the 3-cycle contributes as there are only two non-overlapping 1-cycles (not three) and as there is no 2-cycle in the network. The 3-cycle is $(r_1/|q_0|)\sigma_1(r_3/|q_0|)(-1)(r_2/|q_0|)\sigma_2=-\rho_1\rho_2$.
The Hurwitz determinants, determined from the above principal minors using Eqs.~\ref{eq:routh1}--\ref{eq:routh3}, are:
\begin{align}
\label{eq:jenkin-maxwelldel1} \Delta_1 &=  \rho_1+\rho_2                             \\
\label{eq:jenkin-maxwelldel2} \Delta_2 &= \rho_1\rho_2 (\rho_1+\rho_2-1)   \\
\label{eq:jenkin-maxwelldel3} \Delta_3 &= \rho_1^2\rho_2^2 (\rho_1+\rho_2-1).
\end{align}
One can also derive these determinants directly from the cycles in the graph using Eqs.~\ref{eq:routh1cycle}--\ref{eq:routh3cycle}, with the particular topology of the Jenkin-Maxwell network (two non-overlapping 1-cycles, no 2-cycles, one 3-cycle) leading to the following reduced form:
\begin{flalign}
\Delta_1 &= -c_1\\
\Delta_2 &= -c_1\cdot\overline{c_1c_1}+c_0\cdot c_3\\
\Delta_3 &= c_1\cdot\overline{c_1c_1}\cdot c_3-c_0\cdot c_3\cdot c_3.
\end{flalign}
Upon plugging in for the cycles, the same expressions in Eqs.~\ref{eq:jenkin-maxwelldel1}--\ref{eq:jenkin-maxwelldel3} obtain. The only negative term in $\Delta_2$ is $c_0\cdot c_3=-\rho_1\rho_2$. In $\Delta_3$, the only negative term is $-c_3\cdot c_3=-\rho_1^2\rho_2^2$. These are the critical \textit{multiplicative topologies} present in the influence topology. A multiplicative topology is simply a product of multiple subgraphs (overlapping or non-overlapping) of the network. The notion of multiplicative topology captures much better the true nature of the destabilizing structures in the graph than previous notions of \textit{critical fragments} (which often do not account for the underlying \textit{multiplicative} aspect).  

The stability phase space is displayed in Fig.~\ref{fig:jenkin-maxwell}B. As $\rho_1$ and $\rho_2$ are both assumed positive, the condition for stability can be summarized as:
\begin{equation}
\label{eq:jenkin-maxwellstab}\rho_1+\rho_2-1 > 0,
\end{equation}
or, for $\Psi\equiv\rho_1+\rho_2$ (with $\Psi$ strictly positive), simply $\Psi>1$. This additional reduction of the problem to a single parameter arises from the symmetric contributions of $r_1$ and $r_2$ to the principal minors (swapping of $r_1$ and $r_2$ in Fig.~\ref{fig:jenkin-maxwell}A would lead to the same criterion).
For $\rho_1+\rho_2-1<0$, both $\Delta_2$ and $\Delta_3$ are negative, giving two unstable roots according to the number of sign changes in the Routh array $V(+,+,-,+)$ (Eq.~\ref{eq:routharray}).

For the specific Jenkin-Maxwell network defined by the parameters of Eqs.~\ref{eq:jenkin-maxwellorig1} and \ref{eq:jenkin-maxwellorig2}, $\rho_1=Y^2/(GB)$ and $\rho_2=FY/(GM)$, which, due to the simple linear dependence of the reactions on the species in Eqs.~\ref{eq:jenkin-maxwell1}--\ref{eq:jenkin-maxwell3}, are independent of the exact location of the single steady-state solution, which, at any rate, is located at  $x_1^s=0$, $x_2^s=(P-R-W)/G$, and $x_3^s=V_1+W/F$. Using the above definitions of $\rho_1$ and $\rho_2$, the condition for steady-state stability becomes:
\begin{equation}
\frac{Y^2}{GB} +\frac{FY}{GM} - 1 > 0.
\end{equation}
Upon multiplication by the positive constant $G/B$, this is identical to the stability criterion obtained by Maxwell through explicit solution of the roots of the cubic characteristic polynomial. Such explicit solution is impossible for networks (and their associated characteristic polynomials) that have dimension $n>4$; however, due to the remarkable properties of the Routh-Hurwitz conditions, one could still derive strong topological/algebraic constraints as achieved above through use of the influence topology. The green background in Fig.~\ref{fig:jenkin-maxwell}B indicates that $\rho_1$ and $\rho_2$, according to the definitions above, can assume any positive definite values. For $\rho_1+\rho_2-1>0$, all trajectories converge to the single steady state solution. As $\rho_1+\rho_2-1$ goes from positive to negative, a Hopf bifurcation appears with oscillatory growth to infinity in a particular 2D plane (complex pair of roots with positive real part); convergence to this plane occurs along the third dimension (negative real root). For  $\rho_1+\rho_2-1=0$, oscillations occur in two dimensions with a fixed radius dependent on the initial conditions (pair of purely imaginary roots, similar to the rotation network); convergence to this 2D plane of rotation occurs along the third dimension (negative real root).

It is worth emphasizing that the simple two-parameter condition $\rho_1+\rho_2-1>0$ and the corresponding stability phase space displayed in Fig.~\ref{fig:jenkin-maxwell}B were determined solely from consideration of the influence topology, which is itself completely defined by the graph of nodes and signed directed edges in Fig.~\ref{fig:jenkin-maxwell}A. Aside from the signs of the stoichiometries and monotonicities, no further specification of the exact functional forms of the reactions was required, nor was the number of steady states necessary to specify (only that they should all lie outside the unstable domain displayed in Fig.~\ref{fig:jenkin-maxwell}B for assurance of the network's stability).

\section{van der Pol network} 
\begin{figure}[]
\vspace{-4cm}
\centerline{     \includegraphics*[width=0.5\textwidth]{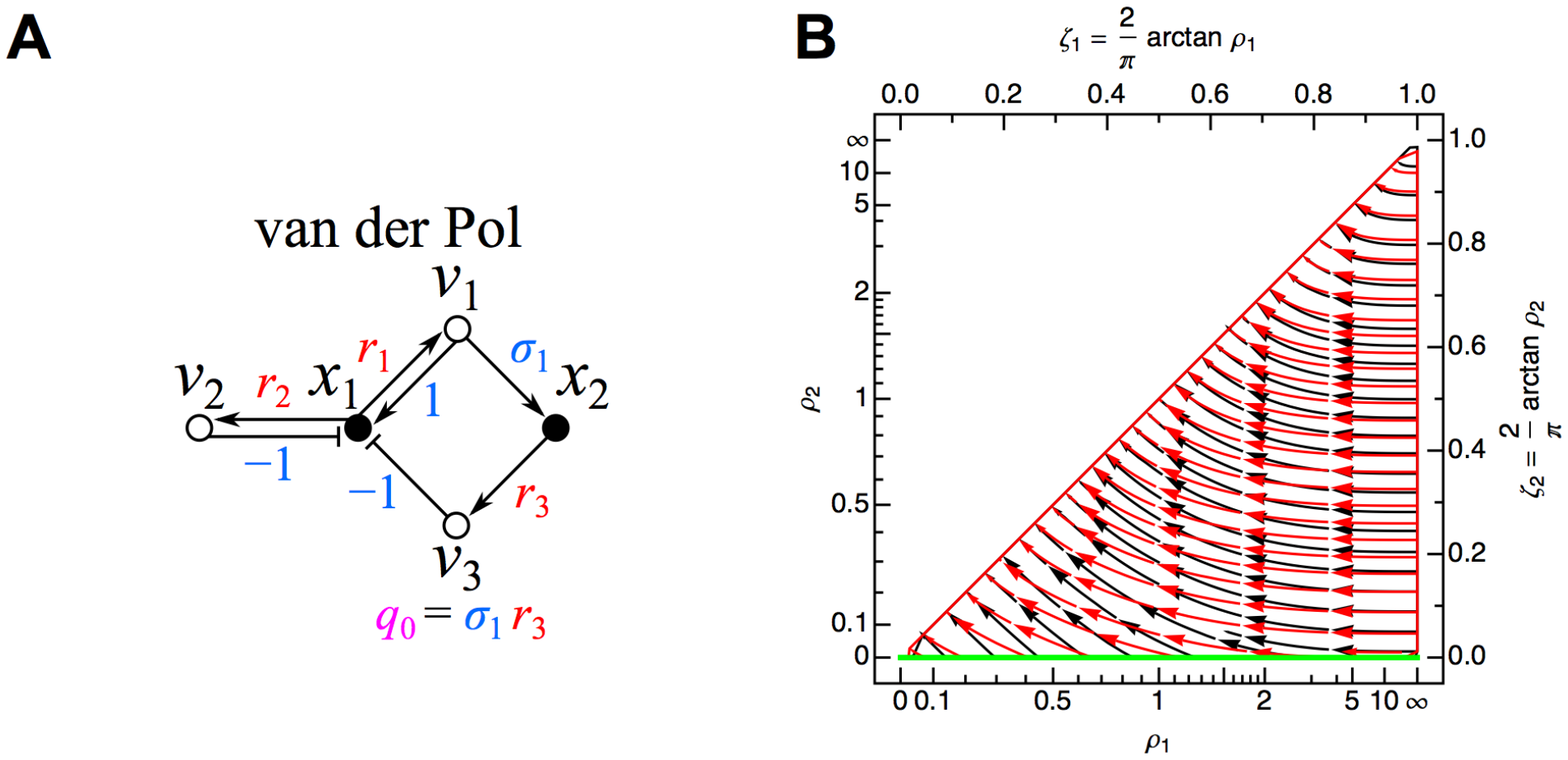}}
\vspace{-4.6cm}
      \caption{Van der Pol network. (A) Influence topology. Cycle compaction allows definition of $q_0=\sigma_1 r_3$. Temporal scaling of all Jacobian edges to $|q_0|$ leaves only $\rho_1=r_1/|q_0|$ and $\rho_2=r_2/|q_0|$. (B) Stability phase space. Flows in the plot map the zones over which $\Delta_1$ (black) and $\Delta_2$ (red) are negative. The unstable zones for $\Delta_1$ (black) and $\Delta_2$ (red) completely overlap; in this region, two unstable eigenvalues obtain according to the Routh array (see text). The green line indicates the possible set of solutions obtainable for the original van der Pol equations (Eqs.~\ref{eq:vanderpol1} and \ref{eq:vanderpol2}).  See Fig.~\ref{fig:jenkin-maxwell} for further details.}
      \label{fig:vanderpol}
\end{figure}

The van der Pol network was first proposed in 1926 \cite{van_der_pol_lxxxviii._1926} as a model for stable oscillations in an electronic circuit:
\begin{equation}
\ddot{x}_1-\mu(1-x_1^2)\dot{x}_1+x_1=0.
\end{equation}
This second-order differential equation can be transformed to the following system of first-order differential equations through use of the Li$\acute{\textrm{e}}$nard transformation \cite{lienard_etude_1928}, $x_2=x_1-x_1^3/3-\dot{x}_1/\mu$, to yield:
\begin{align}
\label{eq:vanderpol1}   \dot{x}_1  = && k_1x_1\hphantom{.}                                              && -k_2 x_1^3          && -k_3x_2                        \\
\label{eq:vanderpol2}   \dot{x}_2  = && \sigma_1k_1x_1.  &&                             &&           
\end{align}
Its generalized form is:
\begin{align}
\dot{x}_1 = && V^1_1\hphantom{,}                        && -V^2_1     && -V^3_2                    \\
\dot{x}_2 =  &&  \sigma_1V^1_1,  &&                     &&           
\end{align}
corresponding to the influence topology displayed in Fig.~\ref{fig:vanderpol}A, with principal minors:
\begin{align}
b_1 &= \rho_1-\rho_2   \\
b_2 &= \rho_1.
\end{align}
For $b_1=c_1$, both 1-cycles in the graph contribute. For $b_2=\overline{c_1c_1}-c_2$, only the 2-cycle $(r_1/|q_0|)\sigma_1(r_3/|q_0|)(-1)=-\rho_1$ contributes as the 1-cycles overlap with each other at the species node (similar overlap at a reaction node would also not be allowed).
The Hurwitz determinants based on the expressions above for the principal minors are:
\begin{align}
\Delta_1 &=  \rho_2-\rho_1  \\
\Delta_2 &= \rho_1(\rho_2-\rho_1),
\end{align}
corresponding to the stability phase space displayed in Fig.~\ref{fig:vanderpol}B.
These can also be derived directly from the cycle-based definitions (where I have already removed terms that are clearly zero based on the influence topology in Fig.~\ref{fig:vanderpol}A):
\begin{flalign}
\Delta_1 &= -c_1\\
\Delta_2 &= c_1\cdot c_2.
\end{flalign}
The corresponding stability phase space is displayed in Fig.~\ref{fig:vanderpol}B.
For $\rho_2-\rho_1<0$, both $\Delta_1$ and $\Delta_2$ are negative, giving two unstable roots according to the number of sign changes in the Routh array $V(+,-,+)$. Using the algebraic redefinition of $\Gamma\equiv\rho_2/\rho_1$ ($\Gamma$ is strictly positive), the condition for instability becomes simply $\Gamma<1$.

For the original van der Pol network defined in Eqs.~\ref{eq:vanderpol1} and \ref{eq:vanderpol2}, the unique steady state solution is $x_1^s=x_2^s=0$ with $\rho_1=k_1/(\sigma_1k_3)$ and $\rho_2=0$ (this complete set of possible solutions is indicated by the green line in Fig.~\ref{fig:vanderpol}B). Since $\rho_2-\rho_1=-k_1/(\sigma_1k_3)<0$, this implies two unstable eigenvalues, which is consistent with the ever-present limit cycle in the network's phase space.

Instead of the Li$\acute{\textrm{e}}$nard transformation, one could alternatively apply the more straightforward transformation of $\dot{x}_1=x_2$, which, upon a final swapping of $x_1$ for $x_2$, leads to:
\begin{align}
\label{eq:vanderpolcan1}   \dot{x}_1  = && k_1x_1\hphantom{.}                                              && -k_2 x_1x_2^2          && -k_3x_2                        \\
\label{eq:vanderpolcan2}   \dot{x}_2  = && \sigma_1k_1x_1.  &&                             &&           
\end{align}
These governing equations are identical to Eqs.~\ref{eq:vanderpol1} and \ref{eq:vanderpol2} aside from the change of $k_2x_1^3\rightarrow k_2x_1x_2^2$ in the second reaction. The generalized form is:
\begin{align}
\dot{x}_1 = && V^1_1\hphantom{.}                        && -V^2_{12}     && -V^3_2                    \\
\dot{x}_2 =  &&  \sigma_1V^1_1.  &&                     &&           
\end{align}
The influence topology is therefore identical to that shown in Fig.~\ref{fig:vanderpol}A aside from a single extra Jacobian arrow from $x_2$ to $v_2$. This extra connection, however, prevents the convenient dimensional reduction obtained for the Li$\acute{\textrm{e}}$nard transformed network, with now three Jacobian parameters and one stoichiometric parameter required to specify the stability phase space instead of the two Jacobian parameters obtained above. This example, therefore, subverts the fundamental nature of the ``topology-then-algebra'' hierarchy implicitly assumed throughout this manuscript. The specific algebraic structure of the governing equations can be critical for specification of the influence topology, with the intriguing possibility --- concretely demonstrated here for the van der Pol network --- that algebraic transformations may exist to convert a given network with a complicated influence topology (requiring many parameters to specify its corresponding stability phase space) to a transformed version having a much simpler influence topology (lower dimensional stability phase space).

\section{Lotka-Volterra network} 
\begin{figure}[]
\vspace{-2cm}
\centerline{     \includegraphics*[width=0.5\textwidth]{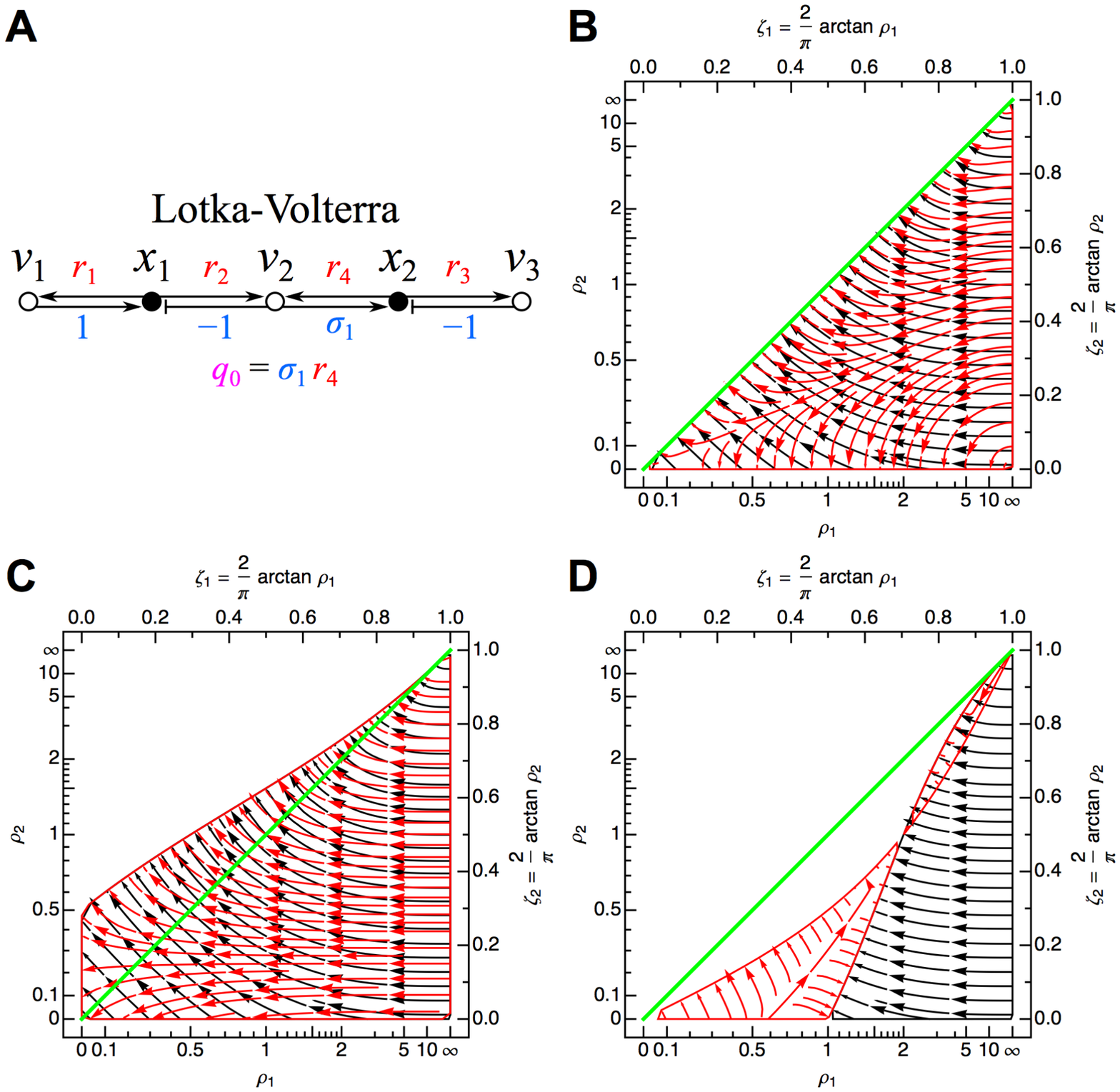}}
\vspace{-2.4cm}
      \caption{Lotka-Volterra network. (A) Influence topology. Cycle compaction allows definition of $q_0=\sigma_1r_4$. Temporal scaling of all Jacobian edges to $|q_0|$ leaves only $\rho_1=r_1/|q_0|$, $\rho_2=r_2/|q_0|$, and $\rho_3=r_3/|q_0|$. The stability phase space is shown for (B) $\rho_3=1$ ($\sigma_1=1$), (C) $\rho_3=1/2$ ($\sigma_1=2$), and (D) $\rho_3=2$ ($\sigma_1=1/2$). Flows in the plot map the zones over which $\Delta_1$ (black) and $\Delta_2$ (red) are negative.  See Fig.~\ref{fig:jenkin-maxwell} for further details.}
      \label{fig:lotka-volterra}
\end{figure}

Lotka in 1920 \cite{lotka_analytical_1920}, and independently Volterra in 1926 \cite{volterra_variazioni_1926,volterra_variations_1928}, introduced the following network for modeling population oscillations:
\begin{align}
\label{eq:lotka-volterra1}    \dot{x}_1  = && k_1x_1  &&  - k_2x_1x_2                    &&                            \\
\label{eq:lotka-volterra2}    \dot{x}_2  = &&               &&    \sigma_1 k_2x_1x_2    && - k_3x_2,
\end{align}
with $\sigma_1=1$ typically assumed.
The generalized Lotka-Volterra network is:
\begin{align}
\label{eq:lotka-volterragen1}   \dot{x}_1  = &&   V^1_1  &&  -  V^2_{12}                  &&                              \\
\label{eq:lotka-volterragen2}   \dot{x}_2  = &&               &&    \sigma_1 V^2_{12}    && - V^3_2, 
\end{align}
corresponding to the influence topology shown in Fig.~\ref{fig:lotka-volterra}A, with 
principal minors (Eq.~\ref{eq:minor}):
\begin{align}
b_1 &= 1+\rho_1 -\rho_2 -\rho_3   \\
b_2 &= \rho_1+\rho_2\rho_3-\rho_1\rho_3,
\end{align}
and Hurwitz determinants:
\begin{align}
\Delta_1 &=  \rho_2+\rho_3-\rho_1-1   \\
\label{eq:LVD2} \Delta_2 &= (\rho_2+\rho_3-\rho_1-1)( \rho_1+\rho_2\rho_3-\rho_1\rho_3).
\end{align}
The above results can also be obtained directly from the cycle product-based expressions of the Hurwitz determinants (Eqs.~\ref{eq:routh1cycle} and \ref{eq:routh2cycle}):
 \begin{flalign}
 \Delta_1&= -c_1\\
\label{eq:LVC2} \Delta_2&= -c_1\cdot\overline{c_1c_1},
\end{flalign}
where I have already removed terms that are clearly zero based on the influence topology. The second term in the above product for $\Delta_2$ corresponds to pairs of non-overlapping 1-cycles ($\overline{c_1c_1}$) of which there are clearly three in the graph of the influence topology (yielding the second group of terms in Eq.~\ref{eq:LVD2}). The stability phase space over $\rho_1$-$\rho_2$ is displayed for different values of $\rho_3$ in Figs.~S\ref{fig:lotka-volterra}B--D.

For the more traditional functional form of the Lotka-Volterra system given in Eqs.~\ref{eq:lotka-volterra1} and \ref{eq:lotka-volterra2}, the unique steady state solution is:
\begin{align}
x_1^s &= k_3/(\sigma_1k_2)\\
x_2^s &= k_1/k_2.
\end{align}
At the steady state, $r_1=k_1$, $r_2=k_1$, $r_3=k_3$, $r_4=k_3$, and $q_0=\sigma_1 r_4 = \sigma_1 k_3$, giving:
\begin{align}
\rho_1 &= \frac{1}{\sigma_1}\frac{k_1}{k_3}   \\
\rho_2 &= \frac{1}{\sigma_1}\frac{k_1}{k_3}    \\
\rho_3 &= \frac{1}{\sigma_1}.
\end{align}
No matter the value of $\sigma_1$, the steady-state solution will always lie on the $\rho_1$-$\rho_2$ diagonal in the stability phase space (green line in Figs.~S\ref{fig:lotka-volterra}B-D). The Hurwitz determinants are:
\begin{align}
\Delta_1 &= \rho_3-1        \\
\Delta_2 &= (\rho_3-1)\rho_1.
\end{align}
For the typically assumed value of $\sigma_1=1$, we obtain $\rho_3=1$ and $\rho_1=\rho_2=k_1/k_3$ and the following critical values for the Routh-Hurwitz conditions: $\Delta_1=0$ and $\Delta_2=0$ (Fig.~\ref{fig:lotka-volterra}B), which prevents any conclusion about the stability of the network at first order. For $\sigma_1=2$, $\rho_3=1/2$ and $\rho_1=\rho_2=k_1/(2k_3)$ with $\Delta_1=-1/2$ and $\Delta_2=-k_1/(4k_3)$, giving two sign changes in the Routh array $V(+,-,+)$ and therefore two eigenvalues with positive real parts (Fig.~\ref{fig:lotka-volterra}C), with the dynamical solutions corresponding to an oscillatory divergence to infinity. For $\sigma_1=1/2$, $\rho_3=2$ and $\rho_1=\rho_2=2k_1/k_3$, with $\Delta_1=1$ and $\Delta_2=2k_1/k_3$ implying a stable network (Fig.~\ref{fig:lotka-volterra}D), characterized by an oscillatory convergence to the steady state solution. 

As already stated above, the reaction functions defined in Eqs.~\ref{eq:lotka-volterra1} and \ref{eq:lotka-volterra2} entail the restriction of the steady state solution to the $\rho_1$-$\rho_2$ diagonal in the stability phase space. The steady state solution can be shifted off the diagonal (even for $\sigma=1$ in Fig.~\ref{fig:lotka-volterra}B) in a way that preserves the influence topology through addition of a constant reaction to one or both of the original governing equations (Eqs.~\ref{eq:lotka-volterra1} and \ref{eq:lotka-volterra2}), or through the introduction of more general reaction functions, for example:
\begin{align}
\hspace{-0.1cm} \dot{x}_1    = &&   k_1\exp{x_1}  \hspace{-0.1cm}   && - k_2\sqrt{x_1}\arctan{x_2}         \hspace{-0.1cm}           &&   \hspace{-0.1cm}              \\
\hspace{-0.1cm} \dot{x}_2    = &&                        \hspace{-0.1cm}     && \sigma_1 k_2\sqrt{x_1}\arctan{x_2}    \hspace{-0.1cm}   && - k_3\log{(1+x_2), \hspace{-0.1cm} }
\end{align}
The assumption of other functional forms for the reactions might additionally allow for the existence of more than one steady state solution.

\section{Brusselator network}

\begin{figure}[]
\vspace{-2cm}
\centerline{\includegraphics*[width=0.5\textwidth]{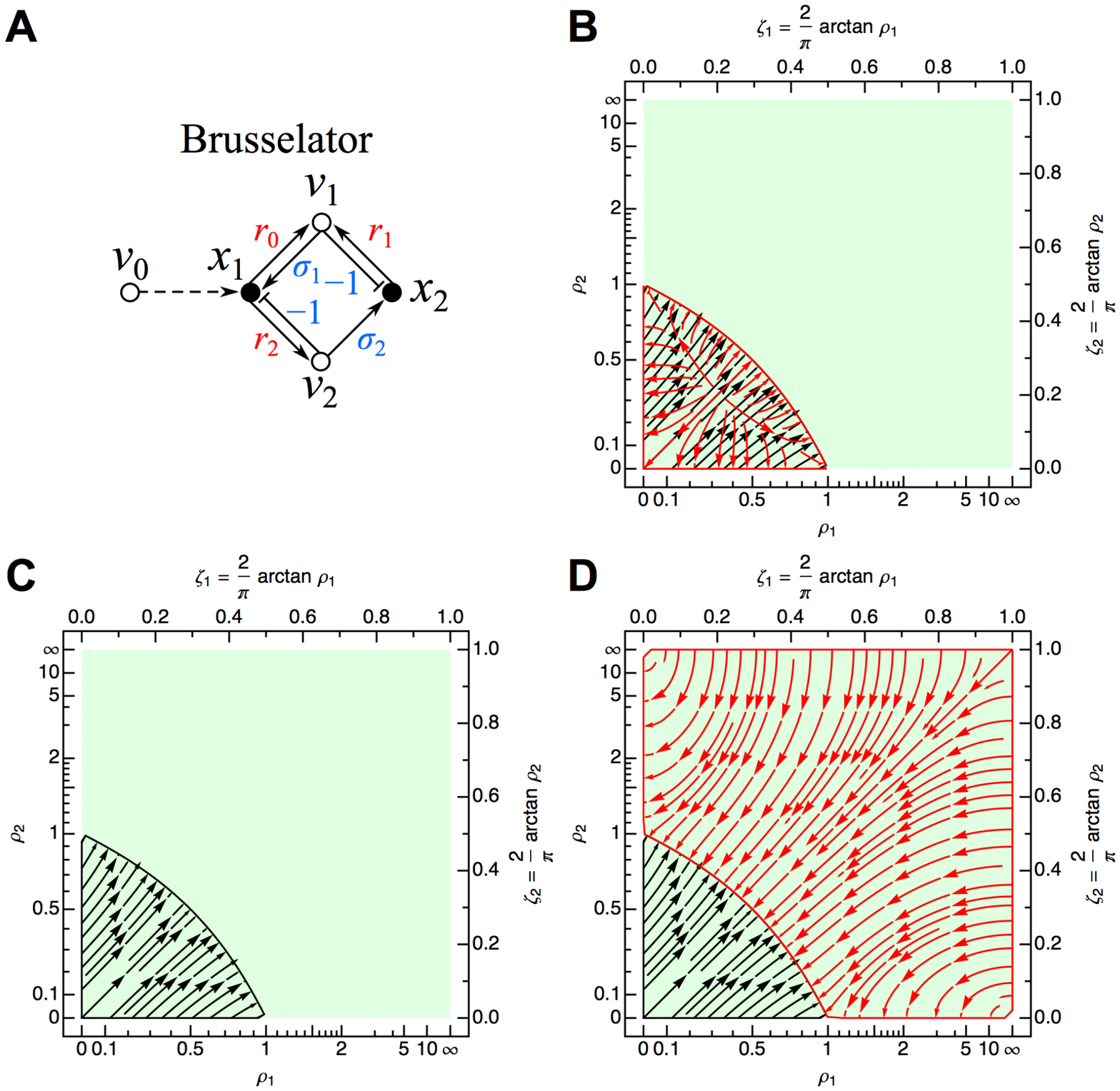}}
\vspace{-2.4cm}
      \caption{Brusselator network. (A) Influence topology. Temporal scaling of all Jacobian edges to $|r_0|$ gives $\rho_1=r_1/|r_0|$ and $\rho_2=r_2/|r_0|$; $\sigma_1$ and $\sigma_2$ must be specified as well. The stability phase space is shown for $\sigma_1=1$ and the following values for $\sigma_2$: (B) $\sigma_2=1/2$, (C) $\sigma_2=1$, and (D) $\sigma_2=2$. Flows in the plot map the zones over which $\Delta_1$ (black) and $\Delta_2$ (red) are negative. See Fig.~\ref{fig:jenkin-maxwell} for further details.}
      \label{fig:brusselator}
\end{figure}

The Brusselator was proposed by Prigogine \& Lefever in 1968 \cite{prigogine_symmetry_1968} to account for oscillations in the Belousov-Zhabotinsky reaction \cite{belousov_periodically_1959,zhabotinsky_periodical_1964}:
\begin{align}
\dot{x}_1  = && k_0                          && + \sigma_1k_1x_1^2x_2                 &&  -k_2x_1\hphantom{,}                   \\
\dot{x}_2  = &&                                 && -k_1x_1^2x_2                                 && +\sigma_2k_2x_1,
\end{align}
with $\sigma_1=1$ and $\sigma_2<1$ (by definition of the original Brusselator network).
Its generalized version is:
\begin{align}
\dot{x}_1  = &&   V^0                              &&  +  \sigma_1 V^1_{12}    && -V^2_1             \hphantom{,}                \\
\dot{x}_2  = &&                                       &&    - V^1_{12}                    && +\sigma_2V^2_1,
\end{align}
corresponding to the influence topology shown in Fig.~\ref{fig:brusselator}A, with principal minors:
\begin{align}
b_1 &= \sigma_1 - \rho_1 - \rho_2 \\
b_2 &= \rho_1\rho_2 -   \sigma_1\sigma_2\rho_1\rho_2,
\end{align}
and Hurwitz determinants:
\begin{align}
\Delta_1 &=    \rho_1 + \rho_2-\sigma_1 \\
\Delta_2 &= ( \rho_1 + \rho_2 -\sigma_1 ) \rho_1\rho_2(1 -   \sigma_1\sigma_2).
\end{align}
These expressions can also be obtained directly from the cycle-based defitions of Eqs.~\ref{eq:routh1cycle} and \ref{eq:routh2cycle} (where I have already removed terms that are zero);
\begin{flalign}
   \Delta_1&= -c_1\\
\Delta_2&= -c_1\cdot\overline{c_1c_1} +c_1\cdot c_2.
\end{flalign}
The stability phase space is displayed in Figs.~\ref{fig:brusselator}B--D for $\sigma_1=1$ and different values of $\sigma_2$. If $1-\sigma_1\sigma_2$ is positive, then instability can only occur for $\rho_1+\rho_2-\sigma_1<0$ (Fig.~\ref{fig:brusselator}B). Defining $\Psi\equiv(\rho_1+\rho_2)$ (with $\Psi$ strictly positive), this amounts to $\Psi<\sigma_1$ for instability, or, upon the further algebraic redefintion $\Gamma\equiv\Psi/\sigma_1$ (with $\Gamma$ strictly positive), this becomes simply $\Gamma<1$.

For the original Brusselator, $\sigma_1=1$ and $\sigma_2=(k_2-a)/k_2$ (or $1-\sigma_2=a/k_2$). Both $\sigma_2$ and $k_2$ are assumed greater than zero, implying the further restrictions of $k_2>a$ and $0<\sigma_2<1$. These definitions give:
\begin{align}
\Delta_1 &=   \rho_1 + \rho_2-1 \\
\Delta_2 &=  ( \rho_1 + \rho_2 -1 ) \rho_1\rho_2\frac{a}{k_2}.
\end{align}
For $\rho_1+\rho_2-1<0$, both $\Delta_1$ and $\Delta_2$ are simultaneously negative, corresponding to a transition from zero to two unstable eigenvalues and therefore the possibility of a Hopf bifurcation and limit cycle. 
\noindent The single steady state for the Brusselator is located at:
\begin{align}
x_1^s&=k_0\\
x_2^s&=\frac{k_2-a}{k_0 k_1},
\end{align}
giving
\begin{align}
\rho_1&=\frac{k_0^2k_1}{2(k_2-a)}\\
\rho_2&=\frac{k_2}{2(k_2-a)}.
\end{align}
Upon appropriate choices for $a$, $k_0$, $k_1$, and $k_2$, both $\rho_1$ and $\rho_2$ can assume any value in the stability phase space (green region of Figs.~\ref{fig:brusselator}B--D). The condition for instability to obtain, $\rho_1+\rho_2-1<0$, becomes:
\begin{equation}
\frac{k_0^2k_1}{a} - \frac{k_2}{a} + 2 < 0.
\end{equation}
Defining $A^2\equiv k_0^2k_1/a$ and $B \equiv k_2/a-1 $, this yields the standard result of $B>1+A^2$.

For $1-\sigma_1\sigma_2=0$, $\Delta_2=0$, implying a reduction in dimensionality and therefore only a single real eigenvalue with sign opposite to that of $\Delta_1$ (Fig.~\ref{fig:brusselator}C). For $1-\sigma_1\sigma_2<0$, 
$\mathrm{sign}(\Delta_2)=-\mathrm{sign}(\Delta_1)$,  which for $\Delta_1\neq0$ will always generate one stable and one unstable eigenvalue according to the one sign change in the Routh array (either $V(+,-,-)$ or $V(+,+,-)$) (Fig.~\ref{fig:brusselator}D).

\begin{figure}[]
\vspace{-2cm}
\centerline{\includegraphics*[width=0.5\textwidth]{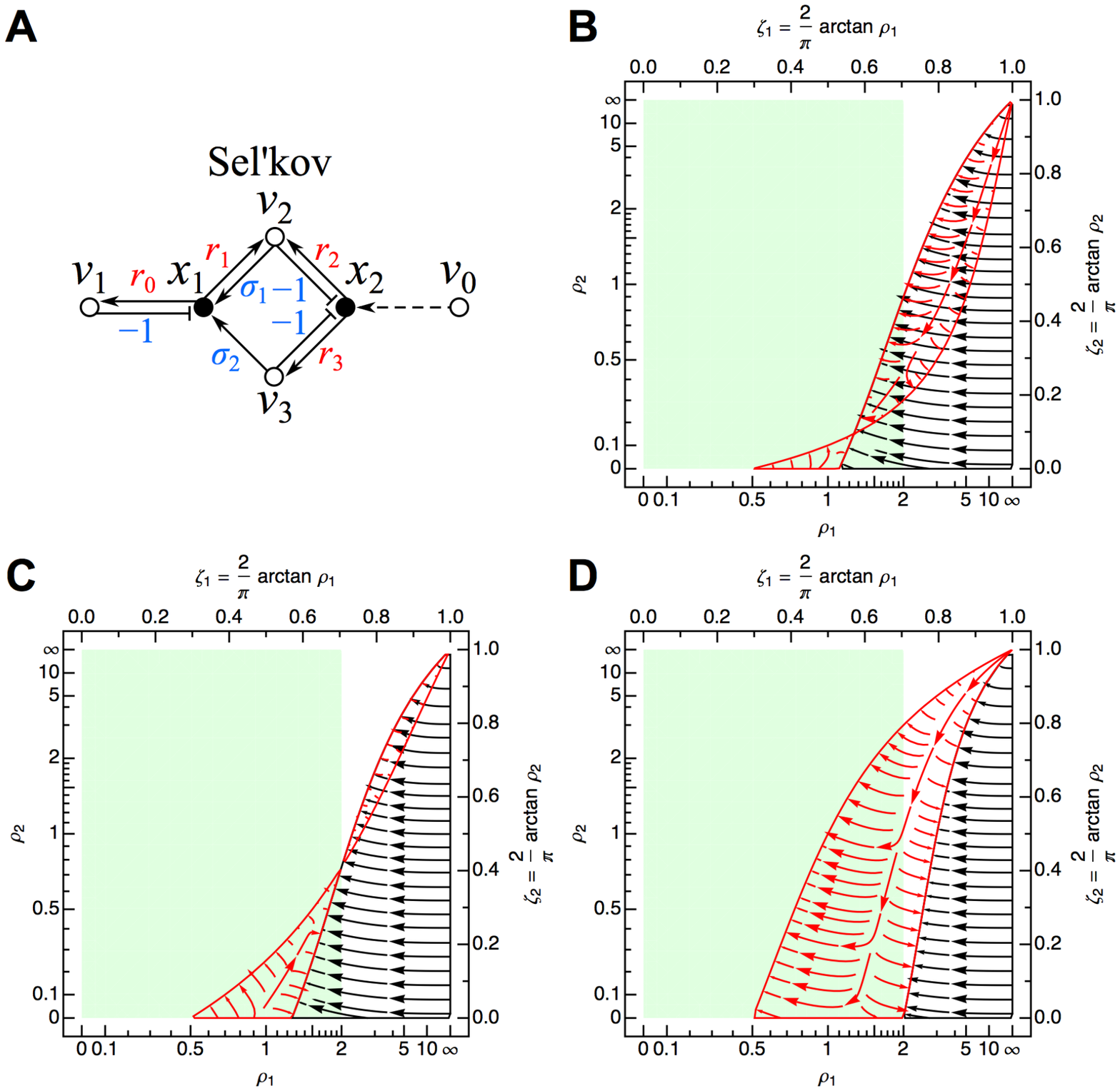}}
\vspace{-2.4cm}
      \caption{Sel'kov network. (A) Influence topology. Temporal scaling of all Jacobian edges to $|r_0|$ gives the parameters $\rho_1=r_1/|r_0|$, $\rho_2=r_2/|r_0|$, and $\rho_3=r_3/|r_0|$. The positive stoichiometric terms $\sigma_1$ and $\sigma_2$ must also be specified independently.  The stability phase space is shown for $\sigma_1=\sigma_2=1$ and (B) $\rho_3=1/10$; (C) $\rho_3=1/4$; and (D) $\rho_3=1$. Flows in the plot map the zones over which $\Delta_1$ (black) and $\Delta_2$ (red) are negative. See Fig.~\ref{fig:jenkin-maxwell} for further details.}
      \label{fig:selkov}
\end{figure}

\section{Sel'kov network} Sel'kov in 1968 \cite{selkov_self_1968} proposed the following simple model to account for glycolytic oscillations:
\begin{align}
\label{eq:selkov1}  \dot{x}_1  = &&               && -k_1x_1   && + \sigma_1k_2x_1^2x_2    &&  +\sigma_2k_3x_2\hphantom{,}              \\
\label{eq:selkov2}  \dot{x}_2  = &&   k_0      &&                 &&  - k_2x_1^2x_2                  &&   - k_3x_2,
\end{align}
with $\sigma_1$ and $\sigma_2$ equal to 1.
The generalized version is:
\begin{align}
\dot{x}_1  = &&               && -V^1_1         && + \sigma_1V^2_{12}    &&  +\sigma_2V^3_2  \hphantom{,}            \\
\dot{x}_2  = &&   V^0      &&                     &&  - V^2_{12}                  &&   - V^3_2.
\end{align}
The influence topology is shown in Fig.~\ref{fig:selkov}A, with principal minors:
\begin{align}
b_1 &= -1 + \sigma_1 \rho_1 -\rho_2 -\rho_3 \\
b_2 &= \rho_2 +\rho_3 - \sigma_1\rho_1\rho_3  -   \sigma_2\rho_1\rho_3,
\end{align}
and Hurwitz determinants:
\begin{align}
\Delta_1 &=  1 + \rho_2 + \rho_3  - \sigma_1 \rho_1  \\
\Delta_2 &=  (1 + \rho_2 + \rho_3  - \sigma_1 \rho_1) \times\notag\\
&\hspace{8em}(\rho_2 +\rho_3 -\sigma_1 \rho_1\rho_3  -\sigma_2 \rho_1\rho_3).
\end{align}
This result can also be obtained directly from the cycle-based definitions (where I have already removed terms that are zero):
\begin{flalign}
   \Delta_1&= -c_1\\
    \Delta_2&= -c_1\cdot\overline{c_1c_1} +c_1\cdot c_2.
\end{flalign}
The corresponding stability phase space is displayed in Figs.~\ref{fig:selkov}B--D for $\sigma_1=\sigma_2=1$ and different values of $\rho_3$.

For the original network (Eqs.~\ref{eq:selkov1} and \ref{eq:selkov2}) with $\sigma_1=\sigma_2=1$, the Hurwitz determinants simplify to:
\begin{align}
\Delta_1 &=  1 + \rho_2 + \rho_3  -  \rho_1  \\
\Delta_2 &=  (1 + \rho_2 + \rho_3  - \rho_1) \left(\rho_2 +\rho_3 - 2\rho_1\rho_3\right).
\end{align}
The steady state solution is:
\begin{align}
x^s_1 &=  \frac{k_0}{k_1} \\
x^s_2 &=  \left(\frac{k_0k_2}{k_1^2}+\frac{k_3}{k_0}\right)^{-1},
\end{align}
at which
\begin{align}
\rho_1 &= 2\left(1+\frac{k_1^2k_3}{k_0^2k_2}\right)^{-1}   \\
\rho_2 &= \frac{k_0^2k_2}{k_1^3} 		\\
\rho_3 &= \frac{k_3}{k_1}. 		
\end{align}
The Sel'kov stability phase space shows interesting structure for three different values of $\rho_3$. A Hopf bifurcation and limit cycle are only possible if $\rho_3<1/4$ and $\rho_1$ and $\rho_2$ map the network to the zone in which both $\Delta_1$ and $\Delta_2$ are less than zero (small black/red overlap region in Fig.~\ref{fig:selkov}B).

\section{Repressilator network}

\begin{figure}[]
\vspace{-4.2cm}
\centerline{\includegraphics*[width=0.5\textwidth]{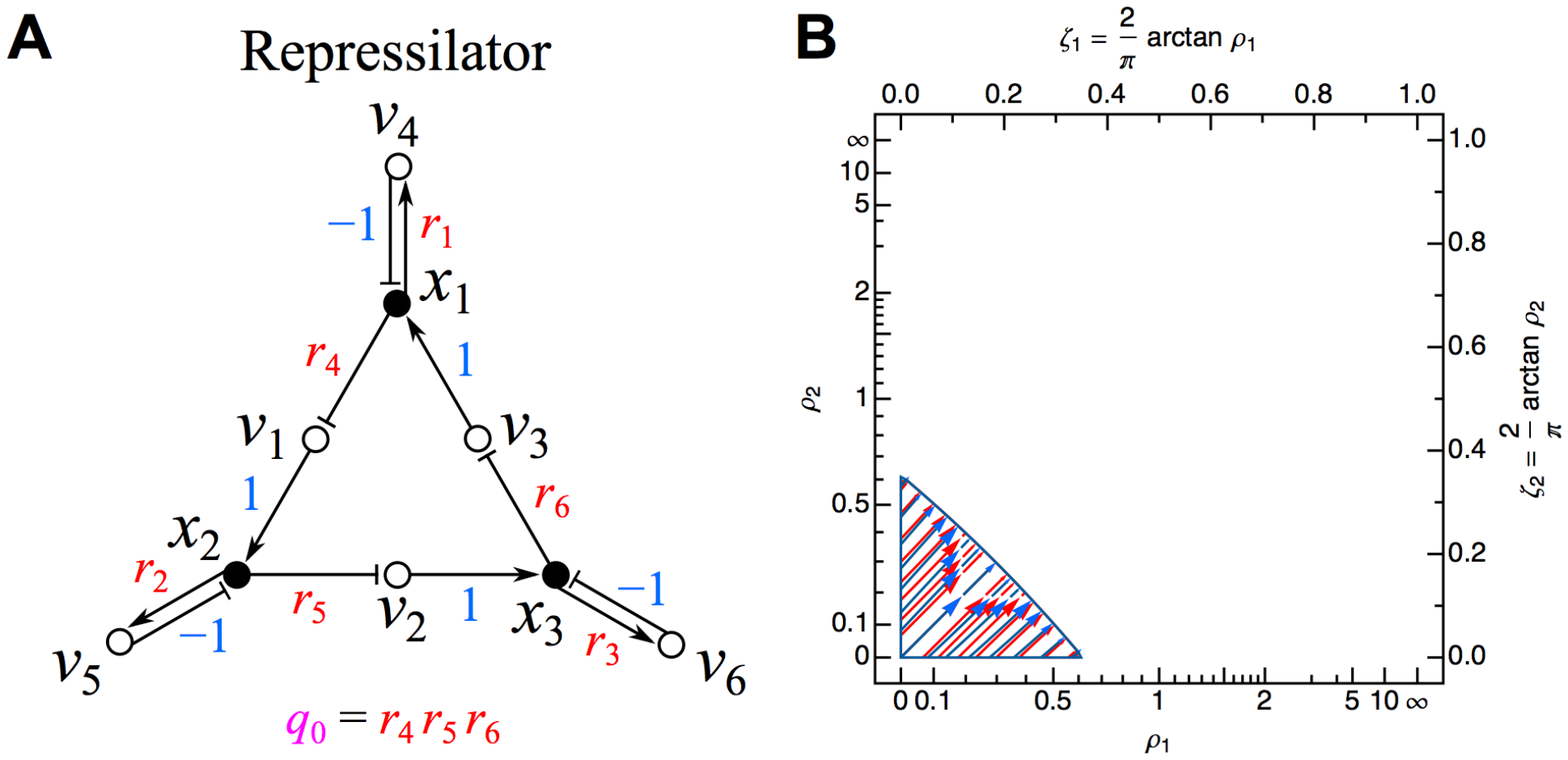}}
\vspace{-4.4cm}
      \caption{Repressilator network. (A) Influence topology. Cycle compaction leads to definition of $q_0=r_4r_5r_6$, which is further removed after temporal scaling, leaving only $\rho_1=r_1/|q_0|^{1/3}$, $\rho_2=r_2/|q_0|^{1/3}$, and $\rho_3=r_3/|q_0|^{1/3}$. (B) Stability phase space for $\rho_3=1$. Flows in the plot map the zones over which $\Delta_1$ (black), $\Delta_2$ (red), and $\Delta_3$ (blue) are negative. For $\rho_3<1$, the domain of instability will be increased towards the upper right (and oppositely for larger $\rho_3$). No green zone is indicated here due to the dependence of the exact steady state solution(s) on the choice of Hill coefficients in the definition of the original network in Eqs.~\ref{eq:repressilator1}--\ref{eq:repressilator3}. See Fig.~\ref{fig:jenkin-maxwell} for further details.}
      \label{fig:repressilator}
\end{figure}

Several different genetic oscillators have been investigated since the first proposal of Monod and Jacob in 1961 \cite{monod_teleonomic_1961}. One particularly famous example is the Repressilator \cite{goodwin_oscillatory_1965,banks_stability_1978,smith_oscillations_1987,elowitz_synthetic_2000}:
\begin{align}
\label{eq:repressilator1}  \hspace{-0.0em} \dot{x}_1  = &&                                      &&                                      &&  \hspace{-1.1em} k_3\frac{1}{1+x_3^{c}}  && \hspace{-0.9em} -k_4x_1        &&                &&        \\
\label{eq:repressilator2}  \hspace{-0.0em} \dot{x}_2  = && \hspace{-0.8em} k_1\frac{1}{1+x_1^{a}} &&                                      &&                                       &&                      && \hspace{-1.1em} -k_5x_2  &&        \\
\label{eq:repressilator3} \hspace{-0.0em} \dot{x}_3  = &&                                      && \hspace{-1.1em} k_2\frac{1}{1+x_2^{b}} &&                                       &&                       &&               &&  \hspace{-1.1em}-k_6x_3,
\end{align}
which has the following generalized form (Fig.~\ref{fig:repressilator}A):
\begin{align}
\label{eq:repressilatorgen1}    \dot{x}_1   = &&                               &&                                &&  V^3_{\overline{3}}   && -V^4_1   &&                &&        \\
\label{eq:repressilatorgen2}    \dot{x}_2   = && V^1_{\overline{1}}  &&                                &&                                 &&               &&  -V^5_2   &&        \\
\label{eq:repressilatorgen3}    \dot{x}_3   = &&                                && V^2_{\overline{2}}   &&                                &&               &&                && -V^6_3,
\end{align}
with principal minors:
\begin{align}
b_1 &= -\rho_1 -\rho_2 -\rho_3  \\
b_2 &= \rho_1\rho_2 + \rho_1\rho_3 + \rho_2\rho_3 \\
b_3 &= -\rho_1\rho_2\rho_3 -1,
\end{align}
and Hurwitz determinants:
\begin{align}
\Delta_1 &=   \rho_1 +\rho_2 +\rho_3   \\
\Delta_2 &= (\rho_1 +\rho_2 +\rho_3)(\rho_1\rho_2 + \rho_1\rho_3 + \rho_2\rho_3) -\rho_1\rho_2\rho_3 -1 \\
\Delta_3 &=  (\rho_1\rho_2\rho_3 +1)\Delta_2,
\end{align}
corresponding to the stability phase space displayed in Fig.~\ref{fig:repressilator}B (for $\rho_3=1$).
These expressions can also be obtained directly from the cycle-based forms (where I have already removed terms that are clearly zero based on the influence topology):
\begin{flalign}
   \Delta_1  &= -c_1\\
   \Delta_2  &= -c_1\cdot\overline{c_1c_1}+c_0\cdot\overline{c_1c_1c_1} +c_0\cdot c_3.\\
   \Delta_3  &=  c_1\cdot\overline{c_1c_1}\cdot\overline{c_1c_1c_1}-c_0\cdot \overline{c_1c_1c_1}\cdot\overline{c_1c_1c_1}  \nonumber\\
     &\hphantom{=} +c_1\cdot\overline{c_1c_1}\cdot c_3 -2c_0\cdot\overline{c_1c_1c_1}\cdot c_3 \nonumber\\
     &\hphantom{=} -c_0\cdot c_3\cdot c_3.
\end{flalign}
After some cancellation in $\Delta_2$, instability can be shown to arise for:
\begin{equation}
\rho_1^2\rho_2+\rho_1^2\rho_3+\rho_2^2\rho_1+\rho_2^2\rho_3+\rho_3^2\rho_1+\rho_3^2\rho_2+2\rho_1\rho_2\rho_3<1.
\end{equation}
The transition from positive to negative occurs simulatenously for $\Delta_2$ and $\Delta_3$, implying the simultaneous appearance of two unstable eigenvalues according to the number of sign changes in the Routh array $V(+,+,-,+)$ (necessary condition for a Hopf bifurcation).
The purely positive sum of terms on the left-hand side is symmetric with respect to exchange of the $\rho_i$ (exchanging the $r_i$ in the influence topology has no effect on the cycle definitions or their non-overlapping contributions to the principal minors). Defining $\Psi$ as this left-hand-side quantity ($\Psi$ is therefore strictly positive) amounts to the single parameter condition of $\Psi<1$ for instability.

As the number and positions of the steady state solutions depend sensitively on the Hill coefficients (yielding complicated expressions even for $a=b=c=1$), no simple general expression exists. Whether the complete stability phase space or only a portion is accessible for a given choice of $a$, $b$, and $c$ may not have a simple answer.

If the 3-cycle is positive rather than negative, the Hurwitz determinants are then:
\begin{align}
\Delta_1 &=   \rho_1 +\rho_2 +\rho_3   \\
\Delta_2 &= (\rho_1 +\rho_2 +\rho_3)(\rho_1\rho_2 + \rho_1\rho_3 + \rho_2\rho_3) -\rho_1\rho_2\rho_3 +1 \\
\Delta_3 &=  (\rho_1\rho_2\rho_3 -1)\Delta_2.
\end{align}
The only possibility for instability is now through $\Delta_3$, which will be negative if $\Psi\equiv\rho_1\rho_2\rho_3<1$. In this region of instability, the Routh array is $V(+,+,+,-)$, implying only one unstable eigenvalue. It is important to note that while there is no possibility for a Hopf bifurcation to arise at any steady state solution, this does not by itself rule out the possibility of a limit cycle.

\section{Discussion} 
The many novel results obtained above, along with the complementary findings of Angeli et al. \cite{angeli_combinatorial_2013}, demonstrate the remarkable utility of the influence topology for addressing the stability of networks described by arbitrary autonomous systems of ordinary differential equations.  Significantly, examination of a network's influence topology, which is based only on the signs of the stoichiometries and monotonicities of its reactions, already restricts the spectrum of its dynamical solutions (including information on the precise numbers of unstable eigenvalues possible) without having to determine the exact steady state solution(s). What is perhaps most striking in the above treatment is the dramatic dimensional reduction often possible, with the many different reaction constants that define the original network reduced to only one or a few parameters for analysis of its stability. For example, the nine reaction parameters plus three initial conditions that define the Jenkin-Maxwell network were reduced to two influence topology parameters, which, due to their symmetry, could be further reduced to a stability criterion based only on a single symmetrized parameter. It is important to emphasize the completely topological origin of this dimensional reduction.

Stoichiometry has been heavily emphasized in the past, almost always under the additional assumption of mass action kinetics \cite{feinberg_existence_1995,feinberg_multiple_1995,craciun_understanding_2006,clarke_stability_1980}. For more general networks, however, the stoichiometric coefficients provide only a limited perspective on network stability. As demonstrated above, the notions of stoichiometric scaling and, even more significantly, cycle compaction prove that variables other than the individual stoichiometric terms are often more suitable for examining a given network's stability. Upon cycle compaction, multiple stoichiometric terms often end up being degenerate with themselves or, even more interestingly, with co-compacted Jacobian terms (e.g. see the Jenkin-Maxwell, van der Pol, and Lotka-Volterra influence topologies presented above). 

Approaches for determining the number of steady states have received more attention in the past than methods for testing steady state stability. It should be noted that all of the networks considered in this paper have (or can have, in the case of the Repressilator) only a single steady state solution no matter the values of the parameters that define the network. Whether this single steady state can become unstable and exactly how it becomes unstable (e.g. through a Hopf bifurcation) is then the interesting question, not steady state multiplicity (which, for the networks considered above, is trivial). It should nevertheless be noted that for more general networks that share the same influence topology as the networks considered above (e.g. Lotka-Volterra-like networks), multiple steady state solutions may be possible depending on the exact form that the reaction functions take, but all of these steady states would still have to lie somewhere on the (unchanged) stability phase space. For an in depth analysis of steady state multiplicity using the influence topology, see the work of Banaji \& Craciun \cite{banaji_graph-theoretic_2009,banaji_graph-theoretic_2010}.

As presented above, the influence topology allows generalization of classical networks to a much larger class sharing the same signs of the reaction stoichiometries and their derivatives with respect to each species. For example, the Lotka-Volterra network was generalized to its influence topology graph, which in actuality represents a much larger class of ``Lotka-Volterra-like'' networks. Such generalizations have always played an important role in deepening our understanding of mathematical objects. While studying the set of all possible interaction networks makes no sense, the set of all possible \textit{influence topologies} is denumerable and can therefore be systematically studied. One can imagine writing down this complete set for a certain fixed number of $n$ species and $m$ reactions and then constructing and examining the associated stability phase space for each individual influence topology. 

How to actually go about algorithmically enumerating all possible signed directed bipartite graphs comprised purely of interconnected cycles (and, therefore, \textit{fundamental} influence topologies) for a fixed number of $n$ species and $m$ reactions while avoiding repeats presents a significant challenge. Aside from this issue, even a small number of species and reactions will generate a lengthy list due to the $2^{J+S}$ different possible unique sign assignments for the $J$ Jacobian and $S$ stoichiometric edges for a given directed bipartite graph. However, the following sign degeneracy significantly reduces this list. Consider a single influence topology graph displaying the positive/negative connections among species \{$x_1$,\ldots,$x_n$\} and reactions \{$v_1$,\ldots,$v_m$\}. For the species node $x_1$, we can make the variable substitution $y_1=-x_1$. That this leads to negation of all its associated edges in the influence topology is simple to show. All Jacobian arrows emanating from node $y_1$ are transformed to $\partial v_k/\partial y_1=-\partial v_k/\partial x_1$; all stoichiometric arrows pointing to node $y_1$ are also negated as the governing equation for species 1 is now $\dot{y}_1=-\dot{x}_1=-f_1=-\sum_k v_k s^k_1=\sum_k v_k(-s^k_1)$. Now consider the reaction node substitution $w_1=-v_1$. All Jacobian arrows that point to this reaction are negated, $\partial w_1/\partial x_i=-\partial v_1/\partial x_i$; all stoichiometric arrows that emanate from this reaction will also be negated, as one can see from the governing equations, $\dot{x}_i=f_i=w_1s^1_i + \sum_{k\neq1} v_k s^k_i=v_1(-s^1_i)+ \sum_{k\neq1} v_k s^k_i$. Note that both substitutions merely represent a definitional degree of freedom of the network variables, with no affect on the network dynamics (the influence topology does not \textit{explicitly} depend on the $x_i$ and $v_k$). The presence of this sign degeneracy significantly reduces the list of all possible unique influence topologies. For $n$-cycle networks, this sign degeneracy is consistent with the fact that only two different types of networks are possible: It is easy to show that the directed edges of any $n$ cycle network can be transformed by sign substitutions into either all arrows ($c_n=1$) or a single blunt arrow and the rest arrows ($c_n=-1$).  For more general networks, the development of an efficient (topological?) algorithm to prove that one influence topology is sign-homomorphic to another would clearly be helpful.

In attempting to write down \textit{the} influence topology for the van der Pol network above, we discovered that it depends on the particular algebraic expression of its governing equations. In this case, two different algebraic versions (the Li$\acute{\textrm{e}}$nard-transformed version and a canonically transformed version) led to influence topologies that differed by only a single link, with the simpler network entailing significantly fewer dimensions to describe its complete stability phase space. This immediately suggests the notion of a \textit{minimal} influence topology for a given network with an associated stability phase space having the fewest possible dimensions; it may of course be degenerate with other similarly minimal topologies, though some of this degeneracy could be removed by the adoption of further criteria such as the influence topology with the smallest cycles, least number of cycles, and/or the least number of each type of edges. For the van der Pol network, the simpler topology (and also minimal?) was connected to the more complicated one through addition of a single edge in the influence topology. Whether more complicated influence topologies for a given network can always be constructed from the minimal topology through the addition of edges (and also reaction nodes) is an interesting question, and one that could easily be disproved by counterexample. The development of algorithms that allow one to find the minimal influence topology (or degenerate set of minimal influence topologies) for a given set of governing equations is a highly interesting open problem.

The influence topology should be especially useful for systems biology and for the de novo construction of biological networks in synthetic biology. Detailed information about biological reaction functions beyond the signs of their stoichiometries and monotonicities is often unavailable (e.g. unknown Hill coefficient \cite{hill_possible_1910}, complicated transcriptional promoter regulation) or sometimes interesting to ignore (e.g. for robustness studies \cite{stelling_robustness_2004,kitano_biological_2004}). The principal benefit of the influence topology is the readily accessible constraints it provides on a given network's possible dynamics, revealing the potential to shift a steady state from stability to instability (or vice versa) as well as what types of transitions are possible (e.g. it provides necessary conditions for a Hopf bifurcation). From a synthetic biology perspective, the stabilization or destabilization of a network in a way that preserves its influence topology could, for example, be achieved through modification of a particular reaction's cooperativity (e.g. as assessed by its Hill coefficient) and therefore the steepness of a given reaction Jacobian.

In this manuscript, six distinct topological interpretations of interaction networks were mentioned. Each interpretation has its own scope for addressing specific questions about networks, including their steady state multiplicity and their stability. The very general results obtained above using the influence topology, along with the complementary results of Angeli et al. \cite{angeli_combinatorial_2013}, demonstrate its remarkable utility for addressing the steady state stability of an arbitrary network with important restrictions on its possible bifurcations. The program outlined above for the enumeration and examination of all possible influence topologies for low dimensional networks should significantly deepen our understanding of the connection between network topology and stability, as will deeper exploration of the relationship of the influence topology to other fundamental topological interpretations. The cycle-based dimensional reductions and topological symmetries discussed above provide a novel appreciation of dynamical system dimensionality and symmetry that complements other perspectives developed over the past two decades \cite{golubitsky_nonlinear_2006,gilmore_symmetry_2007}.

\begin{acknowledgments}

I would like to acknowledge Dr. A. Koseska (MPI of Molecular Physiology) for useful discussions. This manuscript was written using \LaTeX{} (RevTeX4.1) and \BibTeX\  within the TeXShop environment.  All computations were performed in Mathematica\textsuperscript{\textregistered}.  Plots were created using Inkscape, Mathematica\textsuperscript{\textregistered}, and IGOR Pro (Wavemetrics, Inc.).

\end{acknowledgments}


\end{document}